\documentclass[a4paper,11pt,final]{article}

\usepackage[english]{babel}
\usepackage[utf8]{inputenc}
\usepackage{graphicx}
\usepackage{algorithm}
\usepackage{algorithmic}
\usepackage{mathtools}
\usepackage{amssymb}
\usepackage{mathptmx}
\usepackage{xcolor}
\usepackage{amsmath}
\usepackage{tabularx}
\usepackage{multirow}
\usepackage[margin=2cm]{geometry}

\usepackage{lastpage}
\usepackage{fancyhdr}
\usepackage{titling}

\usepackage[caption=false,font=footnotesize]{subfig}

\usepackage{doi}
\usepackage[autostyle]{csquotes}
\usepackage[style=authoryear,citestyle=authoryear-comp,maxcitenames=2,maxbibnames=99,backend=biber,backref=true,dashed=false]{biblatex}
\usepackage
	{todonotes}

\title{Extraction and Analysis of Dynamic Conversational Networks from~TV~Series}
\author{\vspace{-0.2cm}Xavier Bost, Vincent Labatut, Serigne Gueye \& Georges Linarès\\
	\vspace{-0.2cm}\small Laboratoire Informatique d'Avignon -- LIA EA 4128, Université d'Avignon, France \\
    \small\texttt{\{firstname.lastname\}@univ-avignon.fr}}

\hypersetup{
    pdftitle={\thetitle},
    bookmarksnumbered=true,bookmarksopen=true,
	unicode=true,colorlinks=true,linktoc=all,
	linkcolor=blue,citecolor=blue,filecolor=blue,urlcolor=blue,
	pdfstartview=FitH
}

\usepackage{enumitem}
\setlist{nolistsep}

\begin{document}
\maketitle

\begin{abstract}
\addcontentsline{toc}{section}{Abstract}
Identifying and characterizing the dynamics of modern \textsc{tv} series subplots is an open problem. One way is to study the underlying social network of interactions between the characters. Standard dynamic network extraction methods rely on temporal integration, either over the whole considered period, or as a sequence of several time-slices. However, they turn out to be inappropriate in the case of \textsc{tv} series, because the scenes shown onscreen alternatively focus on parallel storylines, and do not necessarily respect a traditional chronology. In this article, we introduce \textit{Narrative Smoothing}, a novel network extraction method taking advantage of the plot properties to solve some of their limitations. We apply our method to a corpus of $3$ popular series, and compare it to both standard approaches. Narrative smoothing leads to more relevant observations when it comes to the characterization of the protagonists and their relationships, confirming its appropriateness to model the intertwined storylines constituting the plots.

\vspace{0.3cm}
\noindent \textbf{Keywords:} \textsc{tv} Series, Plot Analysis, Dynamic Social Network.

\vspace{0.3cm}
\noindent \textcolor{red}{\textbf{Cite as:} X. Bost, V. Labatut, S. Gueye \&  G. Linarès. \href{https://link.springer.com/chapter/10.1007/978-3-319-78196-9_3}{\thetitle}, In: Social Network Based Big Data Analysis and Applications, M. Kaya, J. Kawash, S. Khoury \& M. Y. Day (Eds.), Chap.3, Lecture Notes in Social Networks Series, Springer, 2018. doi: \href{https://doi.org/10.1007/978-3-319-78196-9_3}{10.1007/978-3-319-78196-9\_3}}
\end{abstract}

\section{Introduction}
\label{sec:intro}
\textsc{tv} series became increasingly popular these past ten years. As opposed to classical \textsc{tv} series containing standalone episodes with self-contained stories, modern series tend to develop continuous, possibly multiple, storylines spanning several seasons. However, the new season of a series is generally broadcast over a relatively short period: the typical dozen of episodes it contains is usually being aired over a couple of months. In the most extreme case, the whole season is even released at once. Furthermore, modern technologies, like streaming or downloading services, tend to free the viewers from the broadcasting pace, often resulting in an even shorter viewing time (``binge-watching''). In summary, \textit{modern} \textsc{tv} series are highly continuous from a narrative point of view, but are usually watched in quite a discontinuous way: no sooner is the viewer hooked on the plot than he has to wait for almost one year before eventually knowing what comes next.

The main effect of this unavoidable waiting period is to make the viewer forget the plot, especially when complex. Since he fails to remember the major events of the previous season, he needs a comprehensive recap before being able to fully appreciate the new season. Such recaps come in various flavors: textual synopsis of the plot sometimes illustrated by keyframes extracted from the video stream; extractive video summaries of the previous season content, such as the ``official'' recap usually introduced at the beginning of the very first episode of the new season; or even videos of fans reminding, when not commenting, the major narrative events of the previous season. Though quite informative and sometimes enjoyable, such content-oriented summaries of complex plots always rely on a careful human expertise, usually time-consuming. The question is therefore to know how the generation of such summaries can be partially or even fully automated.

To the best of our knowledge, few works in the multimedia processing field have focused on automatically modeling the plot of a movie. In~\parencite{Guha2015}, the authors make use of low-level, stylistic features in order to automatically detect the typical three-act narrative structure of Hollywood full-length movies. Nonetheless, such a style-based approach does not provide any insight into the story content and focuses on a fixed narrative structure that generalizes with difficulty to the complex plots of \textit{modern} \textsc{tv} series. The benefits of Social Networks Analysis (\textsc{sna}) for investigating the plot content of fictional works have recently been emphasized in several articles. Most focus on literary works: dramas~\parencite{Moretti2011a}, novels~\parencite{Agarwal2012}, etc. In the context of multimedia works, \textsc{sna}-based approaches are even more recent and sparser \parencite{Weng2007}, \parencite{Weng2009}, \parencite{Ercolessi2012}. However, these works focus either on full length-movies or on standalone episodes of classical \textsc{tv} series, where character interactions are often well-structured into stable communities. These approaches consequently do not necessarily translate well when applied to \textit{modern} \textsc{tv} series.

In this paper, we present an \textsc{sna}-based method aiming at providing some insight into the complex plots of \textsc{tv} series, while solving certain limitations of the previous works. It takes as an input a description of the characters' verbal interactions, and outputs a conversational network. Our method considers not only standalone episodes or full-length movies with stable and well-defined communities, but the complex plots of \textsc{tv} series, as they evolve over dozens of episodes. In this case, no prior assumption can be made about a stable, static community structure that would remain unchanged in every episode and that the story would only uncover, and we have to deal with evolving relationships, possibly temporarily linked into dynamic communities. In this case, we are left with building the current state of the relationships upon the story itself, which, by focusing alternatively on different characters in successive scenes, prevents us from monitoring instantaneously the full social network underlying the plot. We thus propose to address this problem by smoothing the sequentiality of the narrative, resulting in an instantaneous monitoring of the current state of any relation at some point of the story. 

Our main contributions are the following. In terms of importance, the first is \textit{narrative smoothing}, the method we propose for the extraction of dynamic social networks of characters. The second is the algorithm we introduce to estimate verbal interactions from a sequence of spoken segments. The third is the annotation of a corpus of $109$ \textsc{tv} series episodes from three popular \textsc{tv} shows: \textit{Breaking Bad}, \textit{Game of Thrones}, and \textit{House of Cards}. The fourth is a preliminary evaluation of our framework on these data, and a comparison with existing methods. This article extends our paper \parencite{Bost2016} on several aspects. First, we deepened the background section and the methods description is more detailed, especially the way we estimated verbal interactions between characters. Second, we now present a comprehensive evaluation of the methods that we proposed to handle conversational interactions, and added examples in the application of narrative smoothing to our dataset. 

The rest of the article is organized as follows. In Section~\ref{sec:review}, we review in further details the previous works related to \textsc{sna}-based plot identification. We then describe the method we propose, by first focusing in Section~\ref{sec:interactions} on the way the verbal interactions between characters are estimated, before detailing in Section~\ref{sec:net_construct} the way a dynamic view of the relationships in \textsc{tv} serial plots can be built independently from the narrative pace. In Section~\ref{sec:exp}, we first systematically evaluate the algorithm we use for estimating verbal interactions; then, we illustrate how our tool can be used by applying it to the three \textsc{tv} serials of our corpus, and we compare the obtained results to existing methods.

\section{Previous Works}
\label{sec:review}
In our review, we distinguish between two kinds of works: the first ones consider a static network resulting from the temporal integration over the whole considered period, which we call \textit{complete aggregation}; the second ones extract and study a dynamic network based on a sequence of smaller integration periods called \textit{time-slices}.

\subsection{Complete Aggregation}
\label{sec:cum_net}
The complete aggregation method consists in building a static network, called a \textit{cumulative network}, in which each node represents a character, and each link models the relationship between the two characters it connects, for the whole considered period of time. It is generally obtained by processing each interaction iteratively, and either updating the weight of the link representing the interaction if it is already present in the graph, or inserting a new link if it is not. Thus, such networks can be weighted. They can also be directed, depending on how the interactions are handled. In the end, a cumulative network is a static graph agglomerating all past relationships, whatever their time ordering. They are widely used in the literature when attempting to apply \textsc{sna} for analyzing the plot of fictional works.

In \parencite{Moretti2011a}, the author emphasizes and illustrates the role \textsc{sna} can play to investigate the plot of literary works. First, after building the network of conversational interactions in a play, the plot, as a sequence of acts occurring over time, is frozen in a spatial, static view that exhibits some underlying patterns: for instance, the conversational network of verbal interactions in Shakespeare's \textit{Hamlet} unveil some critical regions, such as the ``region of death'' in which the whole tragedy consists. Furthermore, a network-based definition of the protagonists should prevent scholars from applying binary, simplifying categories when considering the main and secondary characters. Finally, the \textsc{sna}-based notion of community allows to exhibit two distinct spaces in \textit{Hamlet}'s network: a space of legitimacy around Horatio, associated to the modern democratic state, and a space of usurpation around Claudius, related to the old, declining monarchy. The author further illustrates the benefit of \textsc{sna} for literary studies by considering the question of symmetry in Western and Chinese novels, both at the stylistic level and in the social network of interacting characters.

In \parencite{Weng2007} and \parencite{Weng2009}, relying on similar observations, the authors make use of \textsc{sna} to automatically analyze the plot of a movie. The social network of characters (denoted ``RoleNet'') is built as follows. They first manually characterize the scenes by their boundaries and the characters they involve. They then hypothesize an interaction between two characters whenever they both appear within the same scene. The network is obtained by representing characters as nodes and their interactions by links. These links are weighted according to the number of scenes in which they co-appear, resulting in a \textit{cumulative} representation of time. The authors analyze this network through community detection. They apply this approach to so-called ``bilateral movies'', which involve only two major characters, each of them central in his own community. In \parencite{Weng2007}, the \textit{RoleNet} is used for further investigating the plot, by classifying scenes into one of the two storylines constituting a bilateral movie. In \parencite{Weng2009}, an extended version of the network, without any prior assumption about the number of communities involved, is used as a basis for automatically detecting breakpoints in the story: a narrative breakpoint is assumed if the characters involved in successive scenes are socially distant in the network of characters, as accumulated over the whole story.

In~\parencite{Ercolessi2012}, a similar network of interacting speakers is used, among other features, for clustering scenes of two \textsc{tv} series episodes into separate storylines, defined as homogeneous narrative sequences related to major characters. A standard community detection algorithm is applied to the network of speakers, as built upon each episode, before the social similarity between any pair of scenes is computed, as a relevant high-level feature for clustering scenes into substories.

In~\parencite{Suen2013}, the authors investigate the character interaction networks of a large database of $173$ plays and $580$ movie scripts. The interaction between two characters is incremented by $1$ if they are both found to speak within a sliding window of $10$ successive lines in the screenplays or movie scripts. A number of topological measures such as the average clustering coefficient are processed to describe each network. They are then used as features to perform various classification tasks: distinguishing plays and movies, dating the work, rating it, determining its genre, finding its author, etc.

In summary, cumulative networks can be used as a reliable basis for automatically or manually analyzing the plot of fictional works with well-defined communities, as in plays, full-length movies or standalone episodes of classical \textsc{tv} series\footnote{The website \url{http://moviegalaxies.com/} \parencite{Kaminski2012} provides a convenient way of interactively visualizing such cumulative character networks for a database of about 700 movies.}.

Nevertheless, for \textsc{tv} serials with complex, evolving and possibly parallel storylines, such a static approach is not appropriate. Indeed, a cumulative network built over a long period of time, as in \textsc{tv} serials, gets relatively dense and does not enable to extract meaningful information. More specifically, communities in the final agglomerative network undoubtedly always correspond to substories, partially disconnected in the narrative, but the opposite does not generally stand. Some individuals may have been strongly connected to each other at some point of the story, before some of them interact with other people for some time, resulting in a second substory. Once agglomerated in the cumulative network, such changes in the interaction patterns may be obscured. In some extreme cases, distinct narrative sequences may even result in a complete cumulative graph, for instance in the interaction pattern that follows:

\[
s_{12}^{(1)}...s_{12}^{(2)}s_{13}^{(3)}...s_{13}^{(4)}s_{23}^{(5)}...s_{23}^{(6)}
\]

\noindent where $s_{ij}^{(t)}$ denotes the fact that the $i$\textsuperscript{th} and $j$\textsuperscript{th} characters are the only interacting speakers in the $t$\textsuperscript{th} episode. The three consecutive interaction sequences result in a triangular interaction pattern unable to reflect the three corresponding substories.

\subsection{Time-slices}
\label{subsec:time_slice}
Some works attempt to take into account the evolution of the social network of the characters when analyzing the plot of fictional works. In \parencite{Agarwal2012}, the authors emphasize the limitations of the static, cumulative graph when analyzing the centrality of the various characters of the novel \textit{Alice in Wonderland}. A dynamic view of the social network is then introduced, by building successive static views of the network in every chapter, before standard centrality measures are separately computed in each of them and traced over time for some major characters. Each view corresponds to a so-called \textit{time-slice}, or \textit{time-window}.

Though widely used \parencite{Holme2012} when considering the evolution over time of general networks (i.e. not necessarily narrative ones), time-slice networks, as resulting from the differentiation over some time step of the cumulative network, may still be problematic. In \parencite{Clauset2012}, the authors focus on the critical issue of the time-slice duration, called ``snapshot rate''. It must be chosen carefully to allow to capture a sufficient amount of interactions, but not too many, otherwise one may obtain irrelevant network statistics. The authors then describe a way of automatically estimating the natural time-slice for monitoring over time the evolution of a network of daily contacts in a professional context, where the appropriate time-slice is expected to remain constant.

As a smoother alternative to fixed time-windowing, \parencite{Mutton2004} applies temporal decay to past interactions to estimate the current state of relationships between users of \textit{internet relay chats}. The method is then extended to Shakespeare's plays for monitoring over time the evolution of the network of interacting characters. Nonetheless, we are left with the same kind of issue as with time-windowing approaches: as for the time-slice duration, the ideal value for the decay parameter may be tricky to set.

\begin{figure}[!ht]
  \centering
  \vspace{2mm}
  \includegraphics[width=1\textwidth]{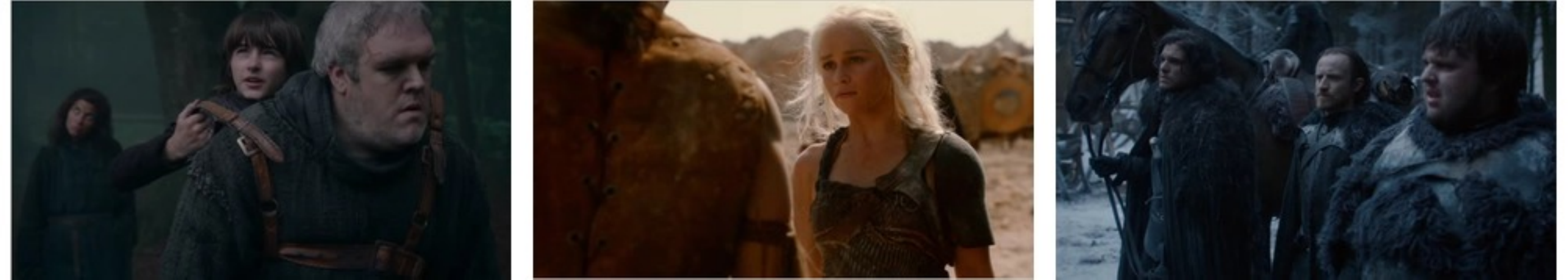}
  \caption{\label{fig:scenes_seq} Three different sets of interacting characters from three consecutive scenes.}
\end{figure}

In order to model the plot of \textsc{tv} serials and allow further analysis, the time-slice should be short enough to capture punctual narrative events related to the social network of characters, but long enough to provide a comprehensive view of the state of the relationships at any point in the story. Unfortunately, getting such a snapshot of the current state of the relationships between the protagonists turns out to be particularly challenging. Unlike the network of physical contacts described in~\parencite{Clauset2012}, the state of the relationships within a story is not fully monitored at any moment, but has to be inferred from the story itself. The narrative usually focuses alternatively on some relationships, possibly belonging to parallel storylines, and only provides a partial view on the network's current state. Some relationships may even take place at the same moment in different places, but will be shown sequentially in successive scenes. Figure~\ref{fig:scenes_seq} illustrates the typical sequential nature of the story as being narrated: three disjoint sets of interacting speakers, possibly at the same time but in different places, are shown sequentially in the story in three successive scenes.

\begin{figure}[!ht]
  \centering
  \vspace{2mm}
  \includegraphics[width=.6\textwidth]{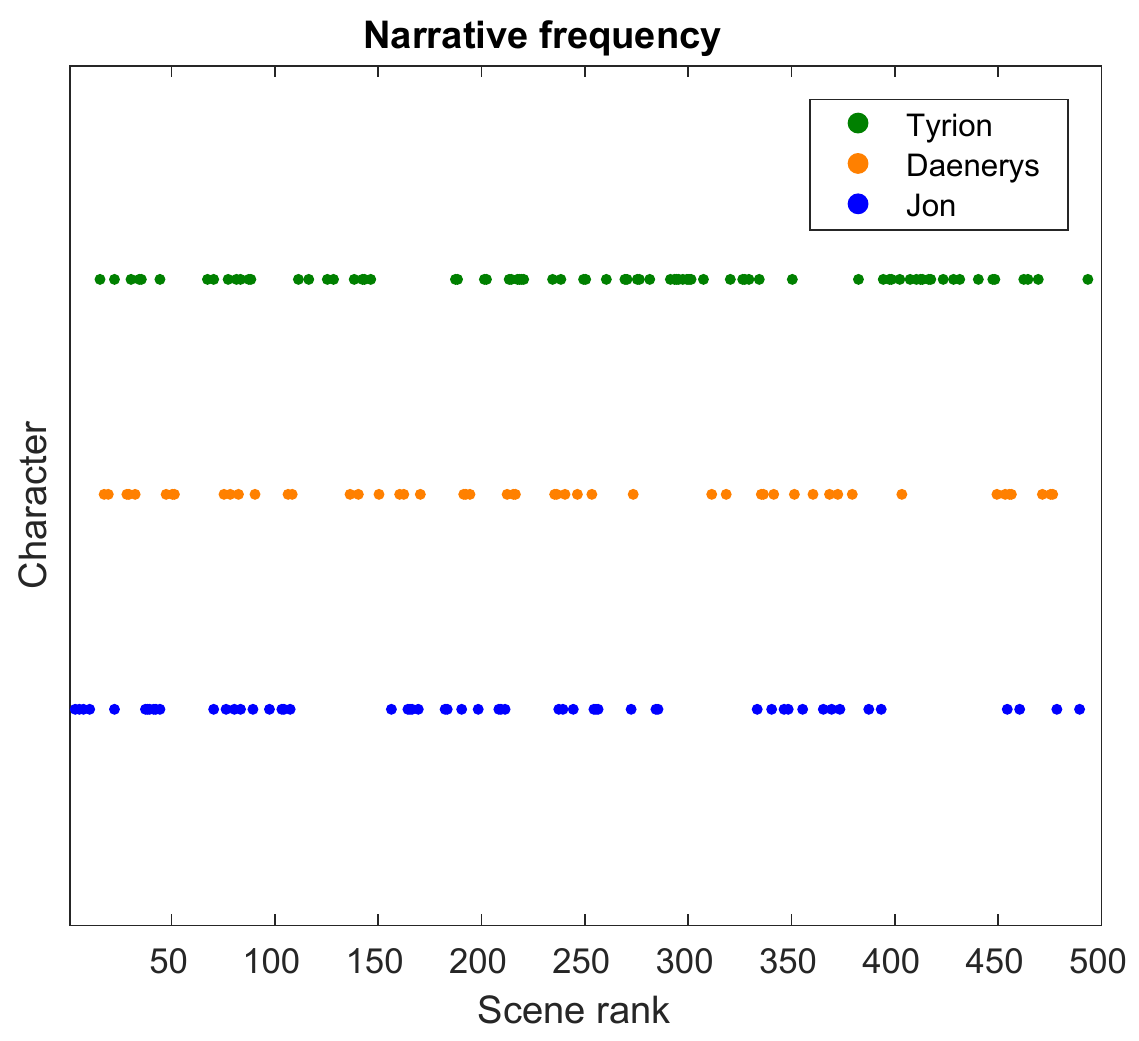}
  \caption{\label{fig:narr_freq} Narrative frequency of three character-based storylines in the first two seasons of \textit{Game of Thrones}.}
\end{figure}

As a consequence, the temporalness of the narrative may be quite different from the temporalness of the underlying network: in particular, the only fact that a group of mutually interacting characters temporarily disappears from the story does not imply that the corresponding characters disappeared from the network. The narrative focus on these interacting characters may only have been postponed by the filmmaker. Furthermore, the pace of activation of the relationships in the story remains largely unpredictable, especially when multiple, disjoint storylines take place in parallel within the narrative. Figure~\ref{fig:narr_freq} plots the scene occurrences of three major character-based storylines in the first two seasons of \textit{Game of Thrones}. Except in the very beginning of the first season, where Jon Snow and Tyrion Lannister meet each other, the three characters interact within well-separated communities.

As can be seen, the way the story alternatively activates these three major storylines does not seem to follow regular patterns. In such a case, the ``ideal'' time-slice may be tricky to set. If too large, it will possibly mask the fast changes usually occurring in the most frequently activated storyline, for instance the story centered around Tyrion. If too narrow, it would lead to irrelevant interpretations of the narrative disappearance of some groups of relationships: the narrative disappearance of Jon Snow's storyline from scene 400 up to scene 450 does definitely not imply that he does not remain socially active in the meantime in his own community. Therefore, the sequential nature of the story should prevent us from identifying the time of the narrative to the time objectively affecting the social network that the story sequentially unveils.

In the rest of this paper, we introduce a novel way of building the dynamic network of interactions between the characters of \textsc{tv} series that allows to fully capture the instantaneous state of every relationship at any point of the story, whatever the pace of activation of each storyline in the narrative.

\section{Estimating Verbal Interactions}
\label{sec:interactions}
\begin{figure}[!ht]
  \centering
  \vspace{2mm}
  \includegraphics[width=1\textwidth]{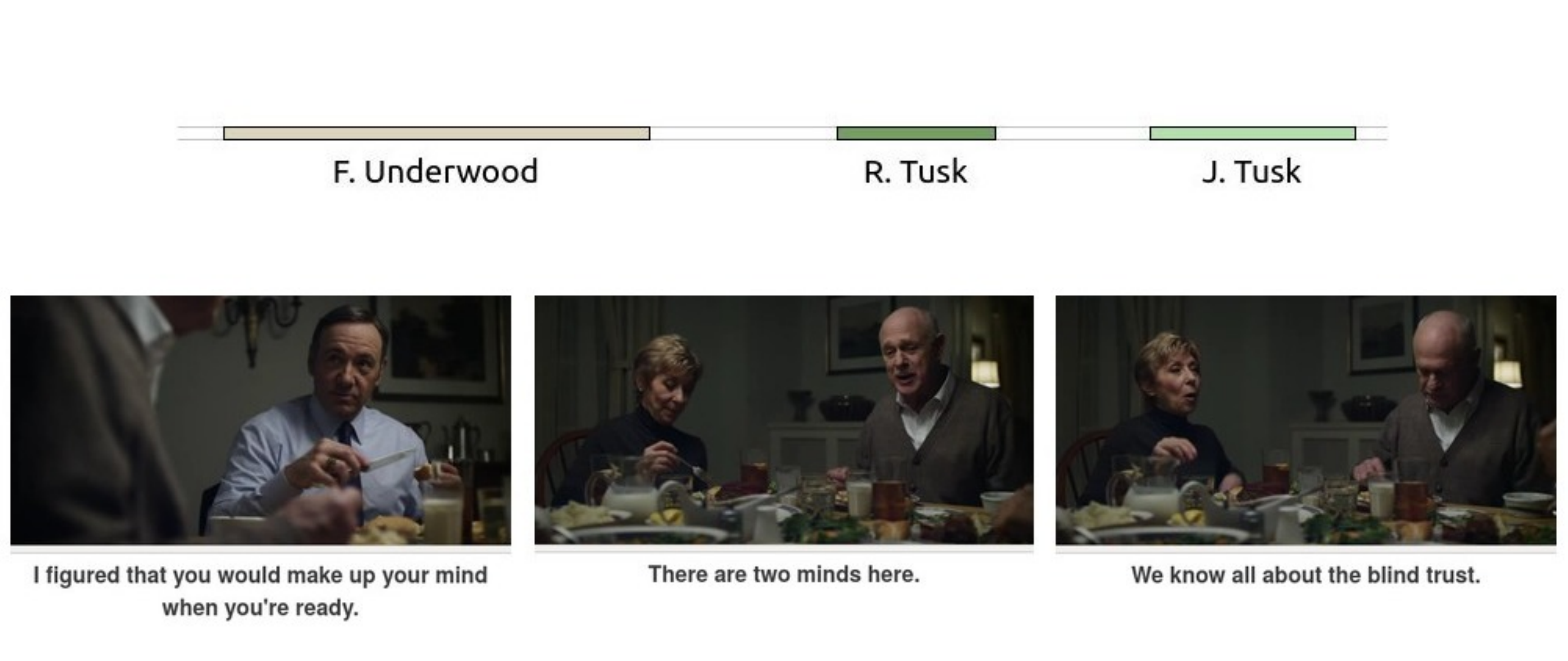}
  \caption{\label{fig:seg_sequence} Sequence of speech segments (top) along with corresponding video frames and subtitles (bottom)}
\end{figure}

Getting an accurate view of verbal interactions within \textsc{tv} serials turns out to be quite challenging, either manually or automatically. When considering a sequence of speech turns within a scene, verbal relationships can be stated as soon as a speaker is talking to an audience, resulting in a directed conversational network, depending on whether someone is talking to, or is being talked by someone. But when a recorded conversation involves more than two speakers, stating who is talking to whom may be tricky. The sequence of speech turns does not convey in itself such an information. By just considering such a sequence, as the excerpt of \textit{House of cards} shown on top of Figure~\ref{fig:seg_sequence}, it is impossible to guess who of the three involved speakers is actually speaking to whom.

Such an information is usually inferred from complementary sources when available, such as the semantic content of the utterances and/or the video recording of the conversation. In this context, an \textit{utterance} is an uninterrupted spoken segment. On Figure~\ref{fig:seg_sequence} the images corresponding to the three utterances, along with their linguistic content, help to disambiguate the verbal situation: from who they are looking at when speaking and/or from the use of personal pronouns, we infer that the first speaker (F.~Underwood) is clearly speaking to the second one (R.~Tusk), while the third one (J.~Tusk) is speaking to the first one. However, guessing to whom the second character is specifically talking remains tricky, even from a careful human expertise using this additional information.

In order to avoid such a tedious human expertise, we consider two main options for automatically estimating verbal interactions: the first one classically relies on the co-occurrence of speakers in scenes~; the second one is an original contribution and relies on the sequence of utterances within each scene.

\subsection{Scene Co-occurrence}
Verbal relationships between characters can first be indirectly deduced from their co-appearance within semantically homogeneous units. For \textsc{tv} serials, which tend to develop complex storylines over several seasons, each typically consisting of a dozen of episodes, possible units are seasons, episodes, or scenes. Seasons or even episodes turn out to be too wide units to provide an accurate view of the actual verbal interactions within \textsc{tv} serials. Because of parallel storylines, stating as interlocutors all the characters co-occurring in a single season or even episode would result in many irrelevant interactions. Considering the scene as a unit is much wiser: a scene in a movie is defined as a homogeneous sequence of actions occurring at the same place within a continuous period of time. The characters co-appearing in a single scene are therefore expected to speak with one another.

A second choice has to be made concerning the kind of character appearance within scenes. Many scenes in movies contain passive characters who do not play any role in the plot and are only physically present but may be talked to by others. Verbal involvement turns out to be much more significant. Although non-verbal relationships, denoted as ``observations'' in \parencite{Agarwal2013}, are still possible between main characters, for instance by only thinking of or by looking at someone else, they usually end up showing verbally in movies. So, by ``characters'', we will always mean ``speakers'', and by ``occurrence'' of the character within a scene, we will mean verbal involvement.

When speech turns, as well as scene boundaries, are explicit, as in plays or movie scripts, the verbal interactions estimated from speaker co-occurrence can be deduced in a fully automatic way. But when the play or the movie is only available as a recorded performance, this information has to be retrieved, either automatically or manually. For \textsc{tv} serials in particular, the scripts are not easily available, or contain only unnormalized and partial information provided on the Web by communities of viewers: we are then left with retrieving speakers as well as scene boundaries.

Though much work has been devoted to automatic detection of scene boundaries, the manual annotation of scene boundaries turns out to be quite straightforward and does not require much time (about 10\% of the film real time duration). The reference scene boundaries, as manually annotated, were thus used for estimating interactions from speakers co-occurrence.

Moreover, automatically detecting ``who spoke when'' in a movie is quite challenging: such a task, known in the speech processing field as \textit{speaker diarization} when performed in an unsupervised way, turns out to be especially tricky when applied to \textsc{tv} series, often containing many speakers talking in adverse acoustic conditions (sound effects, background music...). Despite the benefits of multi-modal approaches (see for instance \parencite{Bost2014} and \parencite{Bost2015}), the error rates obtained when applying speaker diarization tools to \textsc{tv} series remain too high (about 50\%) to serve as a reliable basis for building interaction networks. As we said, the speakers are thus manually indicated, by labeling the subtitles according to the corresponding speakers. This annotation step is much more time-consuming than for scene delimitation, requiring in average as much time as the real duration of each film.

\begin{figure}[!ht]
  \centering
  \vspace{2mm}
  \includegraphics[width=1\textwidth]{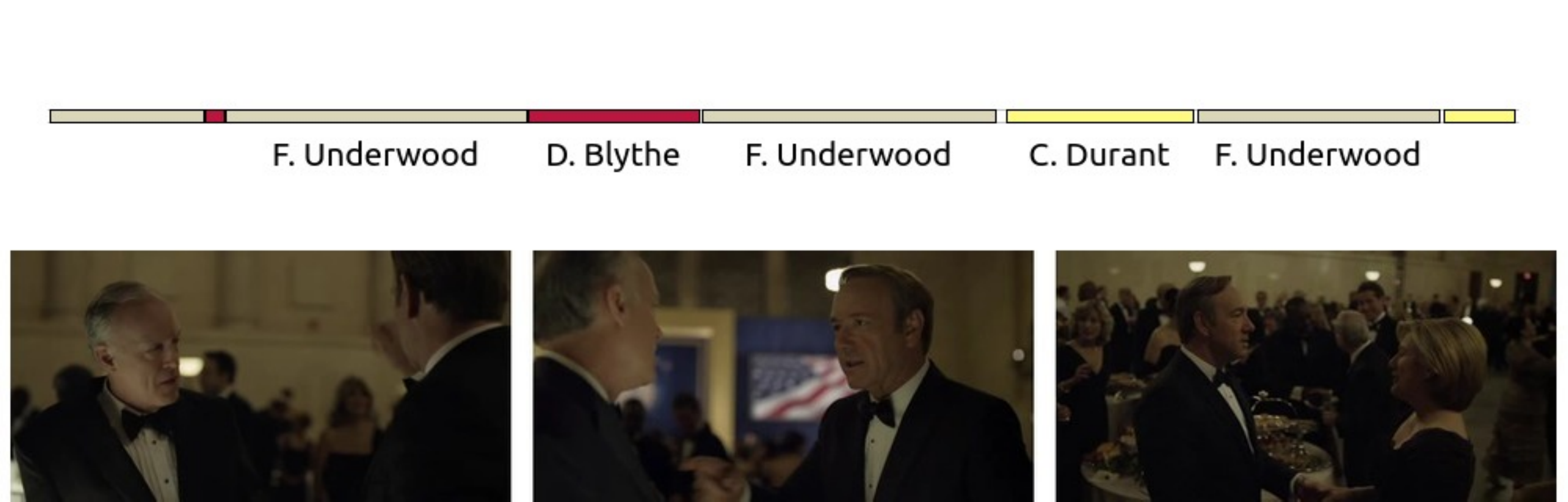}
  \caption{\label{fig:irrel_seq} Two consecutive dialogue sequences within the same scene.}
\end{figure}

Nevertheless, though much more relevant than larger-grained units, the scene used as a way of capturing the verbal interactions between characters may result in weak, sometimes irrelevant, interactions: if being at the same place at the same time is usually required to consider that several persons interact, it is rarely sufficient. Figure~\ref{fig:irrel_seq} shows two consecutive dialogues extracted from the \textsc{tv} serial \textit{House of Cards}, and belonging to the same scene. Three speakers are involved, but without any interaction between the second (\textit{D.~Blythe}) and the third (\textit{C.~Durant}) ones. The first speaker (\textit{F.~Underwood}) is talking to \textit{D.~Blythe} in the first sequence, then is moving to \textit{C.~Durant} and starts discussing with her.

\begin{figure}[!ht]
  \vspace{2mm}
  \centering
  \includegraphics[width=.4\textwidth]{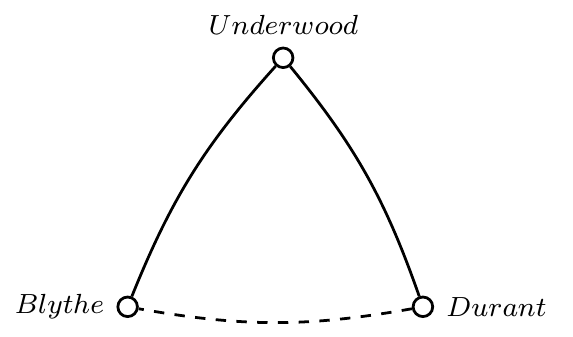}
  \caption{\label{fig:co_occurr} Co-occurrence network based on the sequence shown on Figure~\ref{fig:irrel_seq}. The interaction wrongly introduced is drawn in dash line.}
\end{figure}

The resulting interaction triangle, based on this scene co-occurrence, is shown on Figure~\ref{fig:co_occurr}: \textit{D.~Blythe} and \textit{C.~Durant} are linked whereas they are not involved in any direct verbal interaction. One way of addressing this issue would be to consider even smaller units than scenes, but such a notion of a ``sub-scene'' may be confusing and difficult to define objectively. Another way of facing this problem would be, instead of globally considering the scene unit, to build the verbal interactions upon the sequence of utterances in each scene.

We now focus on relationships defined in a \textit{strong sense}, as based on personal verbal interactions between characters. The resulting network can therefore be considered as a \textit{conversational network}, in contrast to the co-occurrence network of characters described in~\parencite{Weng2007,Weng2009} and used in~\parencite{Ercolessi2012}.

\subsection{Sequential Estimate of Verbal Interactions}
\label{subsec:seq_estimate}
Instead of globally considering the scene unit, we choose to tackle this problem by identifying the verbal interactions from the sequence of speech turns in each scene, once manually labeled according to the corresponding speakers. In order to estimate the verbal interactions from the single sequence of utterances, we define and apply four basic heuristics addressing four possible utterance subsequences. Rules~\textit{(1, 3, 4)} apply to subsequences made of three utterances, while Rule~\textit{(2)} applies to pairs of successive utterances.

\begin{figure}[!ht]
  \center
  \begin{tabular}{cc}
    \subfloat[\textbf{Rule \textit{(1)}}] {
      \includegraphics[width=.4\textwidth]{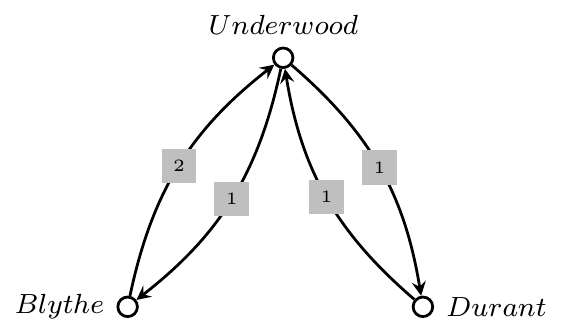}
      \label{subfig:rule_1}
    }
    &
    \subfloat[\textbf{Rule \textit{(2)}}] {
      \includegraphics[width=.4\textwidth]{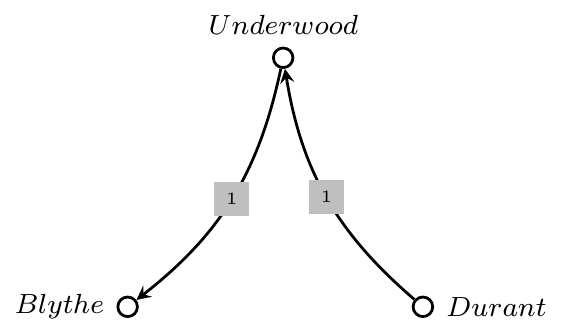}
      \label{subfig:rule_2}
    }
    \\
    \multicolumn{2}{c}{\subfloat[\textbf{Rule \textit{(4)}}] {
      \includegraphics[width=.4\textwidth]{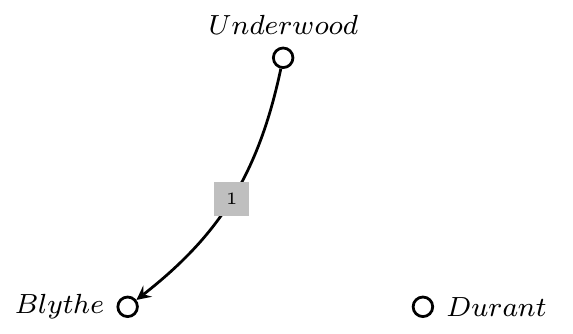}
      \label{subfig:rule_4}
    }}	
    \\	
    \end{tabular}
  \caption{\label{fig:all_rules} Number of directed links resulting from the application of Rules (1--4) to the speech turns sequence shown on Figure~\ref{fig:irrel_seq}.}
\end{figure}

\vspace{0.3cm} \noindent \textbf{Rule~\textit{(1)}~: Surrounded speech turn.} We consider that a speaker $s_2$ is talking to another speaker $s_1$ if this $s_1$ is speaking both after and before $s_2$, resulting in a speech turns sequence $s_1 s_2 s_1$, where each speech turn is labeled according to the corresponding speaker. Subfigure~\ref{subfig:rule_1} shows the subgraph resulting from the application of Rule~\textit{(1)} to the speech turns sequence shown on Figure~\ref{fig:irrel_seq}, where each edge is labeled according to the number of times each speaker is considered as talking to another one.

\vspace{0.3cm} \noindent \textbf{Rule~\textit{(2)}~: Starting and ending speech turns.}  This rule aims at processing the first and last utterances of each sequence $s_1s_2...s_3s_4$ of speech turns, by adding two links $s_1 \rightarrow s_2$ from the first to the second speaker and $s_4 \rightarrow s_3$ from the fourth to the third one. The network resulting from the application of Rule~\textit{(2)} to the sequence of Figure~\ref{fig:irrel_seq} is shown on Subfigure~\ref{subfig:rule_2}.

The last two rules are introduced to process ambiguous sequences of the type $s_1s_2s_3$, where three consecutive speech turns originate in three different speakers: in such cases, the second speaker might be stated as talking to the first one as well as to the third one, or even to both of them. However, such speech turn sequences can often be disambiguated by focusing on speakers involved both before and after the considered sequence.

\vspace{0.3cm} \noindent \textbf{Rule~\textit{(3)}~: Local disambiguation.} We distinguish two variants of this rule. Rule~\textit{(3a)} applies when the second speaker appears before the sequence, but not after, as in $(s_2)s_1s_2s_3(s_4)$. We then consider $s_2$ is speaking with $s_1$ rather than with $s_3$. Symmetrically, Rule~\textit{(3b)} concerns the case when the second speaker appears after, but not before the sequence, as in $(s_0)s_1s_2s_3(s_2)$, and is therefore assumed to speak to $s_3$.

\vspace{0.3cm} \noindent \textbf{Rule~\textit{(4)}~: Temporal proximity.} When the second speaker is involved in the conversation both before and after the ambiguous sequence, as in $(s_2)s_1s_2s_3(s_2)$, we consider the ambiguous speech turn to be intended for the speaker whose utterance is temporally closer. In the sequence shown on Figure~\ref{fig:irrel_seq}, the fifth, ambiguous utterance would then be hypothesized as intended for the first speaker \textit{D.~Blythe}, resulting in the additional link shown on Subfigure~\ref{subfig:rule_4}. The same Rule~\textit{(4)} is applied when the speaker $s_2$ is not involved in the immediate conversational context.

\begin{figure}[!ht]
  \center
  \begin{tabular}{cc}
    \subfloat[] {
      \includegraphics[width=.4\textwidth]{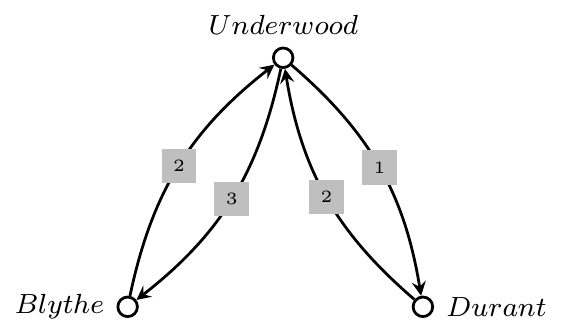}
      \label{subfig:utt_seq_nb}
    }
    &
    \subfloat[] {
      \includegraphics[width=.4\textwidth]{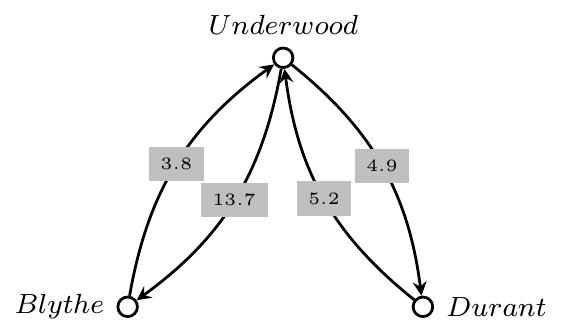}
      \label{subfig:utt_seq_dur}
    }
    \\
    \end{tabular}
  \caption{\label{fig:app_rules} Directed links resulting from the joint application of Rules (1--4) to the speech turns sequence shown on Figure~\ref{fig:irrel_seq}, with weights corresponding either to the number of interactions (\ref{subfig:utt_seq_nb}) or to interaction time in seconds (\ref{subfig:utt_seq_dur}).}
\end{figure}

Figure~\ref{fig:all_rules} and \ref{fig:app_rules} show the directed interactions identified between any two speakers involved in the scene described by Figure~\ref{fig:irrel_seq}, after jointly applying Rules~(1--4). The former considers the rules separately, whereas the latter displays their combined results. On the left-hand part of Figure~\ref{fig:app_rules} (\ref{subfig:utt_seq_nb}), the weight of the links is the number of times one speaker is talking to another one; on its right-hand part (\ref{subfig:utt_seq_dur}), the weight is computed as the total duration of the interaction.

We now describe the algorithm we use to build a dynamic network of interacting speakers able to capture the evolution of the narrative content over time.

\section{Dynamic Conversational Network for Plot Modeling}
\label{sec:net_construct}
The conversational network is directly extracted from the set of verbal interactions identified at the previous stage. In this section, we first present the method we propose to perform this extraction, then illustrate it with a detailed example.

\subsection{Narrative Smoothing}
As stated in Section~\ref{sec:intro}, we would like to get an instantaneous measurement of the intensity of any relationship at any moment, but from the successive partial views of the underlying network that the narrative provides us. Intuitively, a particular relationship may be considered as especially important at some point of the story if the involved characters both speak frequently and a lot to each other: the time interval needed before the reactivation of the interaction in the narrative is expected to be short, and the interaction time is expected to be long whenever the relationship is active in the plot.

\vspace{0.3cm} \noindent \textbf{Preliminary definitions.} Before putting this idea in practice, we need first to introduce several functions, starting with a measure of the interaction between two speakers during a scene. The rule-based method presented in the previous section allows us to estimate who speaks to whom in a given scene. Based on this, it is possible to compute the total duration of speech one speaker directs at another one during the scene, by summing up the durations of all the utterances identified as such. Let us formally identify the scenes by integer, sequential indices. We obtain $h_{ij}^{(t)}$, the total amount of interaction between two speakers $s_i$ and $s_j$ during the $t$\textsuperscript{th} scene, by adding the total duration for which $s_i$ talks to $s_j$, and the total duration for which $s_j$ talks to $s_i$, during this scene. Note that this value is expressed in seconds, and can be zero if the considered speakers do not speak to each other. Moreover, the function $h_{ij}^{(t)}$ is symmetrical relatively to $s_i$ and $s_j$, since we make no distinction between the concerned speakers.

The second measure is the \textit{narrative persistence} of the relationship between speakers $s_i$ and $s_j$ at scene $t$, noted $\Delta_{ij}^{(l)}(t)$. Here, $l$ represents the index of the last scene (relatively to $t$) during which the considered speakers have verbally interacted. This measure is defined as:
\begin{equation}
  \Delta_{ij}^{(l)}(t) = h_{ij}^{(l)} - \sum_{t' = l + 1}^t  \sum_{k \neq i,j} \left ( h_{ik}^{(t')} + h_{jk}^{(t')} \right )
  \label{eq:narr_persist}
\end{equation}
It corresponds to the net balance between the duration $h_{ij}^{(l)}$ of the \textit{last} interaction between the two characters $s_i$ and $s_j$, and the conversational time (represented by the double sum) that $s_i$ and $s_j$ have devoted separately to other characters $s_k$ since then.

The third measure is the \textit{narrative anticipation}, which is defined symmetrically, and noted $\Delta_{ij}^{(n)}(t)$. Here, $n$ denotes the index of the next scene (relatively to $t$) during which the speakers will interact. It is defined as:
\begin{equation}
  \Delta_{ij}^{(n)}(t) = h_{ij}^{(n)} - \sum_{t' = t}^{n-1} \sum_{k \neq i,j} \left ( h_{ik}^{(t')} + h_{jk}^{(t')} \right )
  \label{eq:narr_antic}
\end{equation}
Quite straightforwardly, it is the difference between the duration of the \textit{next} interaction between the considered speakers, and the time they \textit{will} devote separately to other characters in the meantime.

Now, we can start describing the network extraction process itself. Four possible states have to be considered when monitoring a single relationship over time: (1) the relationship is active in the current scene; (2) it has been active in the story and will be active again later; (3) it was active before, but will no longer be active in the narrative; and (4) it has not yet been active in the narrative.

\vspace{0.3cm} \noindent \textbf{(1) Relationship currently active in the story.}
The first case is the simplest one: each time the interaction occurs, its intensity can be estimated in a standard way as the duration of the interaction, expressed in seconds. In any scene $t$ where speakers $s_i$ and $s_j$ are hypothesized as talking to each other, the instantaneous weight $w_{ij}^{(t)}$, which represents the intensity of their relationship, is estimated as follows:
\begin{equation}
  w_{ij}^{(t)} = h_{ij}^{(t)}
  \label{eq:inter}
\end{equation}
where $h_{ij}^{(t)}$ denotes the interaction time, expressed in seconds, between speakers $i$ and $j$ in scene $t$.

The last three cases are much trickier. Between two consecutive occurrences of the same relationship in the story, it would be tempting to consider that the relationship is still (resp. already) active if it is recent (resp. imminent) enough at each moment considered. This is the method adopted in the time-slice framework described in Section~\ref{sec:review}: as long as the relationship is present in the observation window of the network over time, it is supposed active, and inactive as soon as no longer observed. A smoother alternative based on temporal decay is used in \parencite{Mutton2004}.

However, as emphasized in Section~\ref{sec:review}, such a way of handling the past and future occurrences of the relationship is inappropriate for most \textsc{tv} serials. Some interacting characters may be absent from the narrative for an undefined period of time but still be linked in the underlying network, as confirmed by the fact that the last state of the relationship is generally used as a starting point when the characters are re-introduced in the story. Indeed, the temporalness of the narrative should affect a relationship only when at least one of the involved characters interacts with others after and/or before the relationship is active: the relationship between two characters should only get weaker if they interact separately with others before interacting again with one another. This is why, in order to handle the remaining cases, we need to use the previously defined narrative persistence and/or narrative anticipation.

\vspace{0.3cm} \noindent \textbf{(2) Relationship between two narrative occurrences.}
We then define the instantaneous weight $w_{ij}^{(t)}$ of the relationship between the speakers $s_i$ and $s_j$ in any scene $t$ occurring between two consecutive occurrences of their relationship as:
\begin{equation} 
  w_{ij}^{(t)} = \max \left \{ \Delta_{ij}^{(l)}(t), \Delta_{ij}^{(n)}(t) \right \}
\label{eq:between}
\end{equation}

If neither of the two characters speaks to others before they interact again with one another, $w_{ij}^{(t)} = \max \left \{ h_{ij}^{(l)}, h_{ij}^{(n)} \right \}$ and the last (resp. next) occurrence of the relation is considered as still (resp. already) fully present in the network, whatever the number of intermediate scenes the narrative introduces in-between to focus on other plot substories.

\vspace{0.3cm} \noindent \textbf{(3) Relationship after its last narrative occurrence.}
The weight of the relationship between the $i$\textsuperscript{th} and $j$\textsuperscript{th} speakers in any scene $t$ occurring after its very last occurrence in the narrative is expressed as follows, provided that one of the two characters remains involved in the story by interacting with others:
\begin{equation} 
  w_{ij}^{(t)} = \Delta_{ij}^{(l)}(t)
\end{equation}

\vspace{0.3cm} \noindent \textbf{(4) Relationship before its first narrative occurrence.}
Symmetrically, the weight of the relationship between the $i$\textsuperscript{th} and $j$\textsuperscript{th} speakers in any scene $t$ occurring before its first occurrence in the story is computed as follows, as long as one the two characters has already been shown as interacting with other people:
\begin{equation} 
  w_{ij}^{(t)} = \Delta_{ij}^{(n)}(t)
\end{equation}
In the very last case, when neither of the two characters is still (resp. already) active, the weight $w_{ij}$ is set to $-\infty$.

\vspace{0.3cm} \noindent \textbf{Normalization.}
Once the weight $w_{ij}^{(t)}$ has been processed through one of the four possible methods, we use the sigmoid function to normalize it. We note $n_{ij}^{(t)}$ the normalized weight of the relationship between speakers $s_i$ and $s_j$ at scene $t$:
\begin{equation}
  n_{ij}^{(t)} = \frac{1}{1 + e^{-\lambda w_{ij}^{(t)}}}
  \label{eq:weight_norm}
\end{equation}

We chose the sigmoid function to perform such a normalization for two reasons: 1) to get weights ranging from $0$ to $1$, and 2) to model the way the past and future states of a relationship in the narrative could influence its current state at some point $t$. The parameter $\lambda$ is a parameter of sensitivity to the past and future states of the network and was set to $\lambda = 0.01$ in our experiments (high values imply low dependence on the future and past states). This results in an undirected graph $\mathcal{G}^{(t)}$, capturing the instantaneous state of the social network that the story sequentially unveils.

\subsection{Narrative Smoothing Illustrated}
\begin{figure}[!ht]
  \vspace{2mm}
  \centering \includegraphics[width=\textwidth]{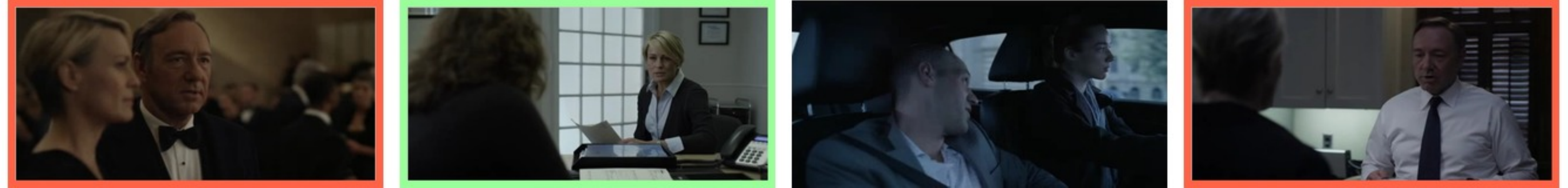}
  \caption{\label{fig:weighting_scheme} Example of application of the weighting scheme to a specific relationship.}
\end{figure}

Figure~\ref{fig:weighting_scheme} shows excerpts of four consecutive scenes in~\textit{House of Cards}, involving five individuals. The first two of them, namely \textit{Francis Underwood} and his wife \textit{Claire}, interact with each other in the first and last scenes (red border) respectively during 30 and 20 seconds, whereas Claire interacts in-between 40 seconds with another person in the second scene (green border) and two other people are talking to one another in the third scene during 50 seconds.

In the first and fourth scenes, Claire and Francis are interacting with each other: according to Equation~\ref{eq:inter}, we then set the weights of their relationship to the corresponding interaction times, respectively $30$ and $20$ seconds.

In the second scene, the last interaction between Claire and Francis is on the one hand weakened by the separate interaction of Claire with someone else during $40$ seconds: the resulting \textit{narrative persistence} of the relationship between Francis and Claire then amounts to $\Delta_{12}^{(1)}(2) = 30 - 40 = -10$ (Equation~\ref{eq:narr_persist}).

On the other hand, the \textit{narrative anticipation} on the next occurrence of the relationship between Francis and Claire then amounts to $\Delta_{12}^{(4)}(2) = 20 - 0 - 40 = -20$ (Equation~\ref{eq:narr_antic}), resulting in an instantaneous weight $w_{12}^{(2)} = \max \{-10, -20\} = -10$ in the second scene.

In the third scene, neither of the two characters is involved: the narrative persistence of their relationship is unchanged, but the narrative anticipation then increases to 20, because no interfering character separates at this point Francis and Claire from their next interaction in the fourth scene. We then have $w_{12}^{(3)} = \max \{-10, 20\} = 20$ and the full resulting sequence of unnormalized, instantaneous weights for the relationship between Claire and Francis is then (30, -10, 20, 20) at the four considered moments.

\section{Experiments and results}
\label{sec:exp}
In this section, we evaluate the whole framework we introduced for building a reliable and informative dynamic social network of interacting characters in \textsc{tv} serials. We first evaluate the basics heuristics we introduced in Subsection~\ref{subsec:seq_estimate} to infer the interacting speakers from the sequence of speech turns, once manually annotated according to the corresponding speakers. We then qualitatively evaluate narrative smoothing, our graph extraction method, by comparing it to both types of methods described in Section~\ref{sec:review}. For this purpose, we focus on every \textsc{tv} serial of our corpus, and explore their plots from the dynamics of their underlying social network of characters. We both analyze the obtained networks from the perspective of the protagonists (nodes) and their relationships (links). We now describe the corpus subset we used when evaluating our methods.

\subsection{Corpus}
\label{subsec:corpus}
Our corpus consists in three very popular \textsc{tv} series: \textit{Breaking Bad} (first 3 seasons), \textit{Game of Thrones} (first 5 seasons), and \textit{House of Cards} (first 2 seasons). We manually annotated the scene boundaries and labeled each subtitle according to the corresponding speaker. The obtained annotations were then used to extract the social networks of characters, by first estimating the verbal interactions according to the rules described in Subsection~\ref{subsec:seq_estimate} and then by using the existing methods presented in Section~\ref{sec:review} as well as our own narrative smoothing approach. The resulting networks are publicly available online\footnote{\url{https://dx.doi.org/10.6084/m9.figshare.2199646}}, along with short videos showing the evolution of the three networks of characters over the seasons considered. Table~\ref{tab:corpus} reports the main features of the first 10 or 11 episodes of each \textsc{tv} series.

\begin{table}[!ht]
  \caption{\label{tab:corpus}{\it Main features of our corpus (first 10/11 episodes).}}
  \centering
  \begin{tabular}{|l||c|c|c|}
    \hline
    \textsc{corpus} & \textsc{BB} & \textsc{GoT} & \textsc{HoC} \\
    \hline
    \hline
    Number of episodes & 11 & 10 & 11 \\
    Total duration (\textsc{h:mm:ss)} & 8:25:04 & 8:28:51 & 8:21:42\\
    Speech coverage (\%) & 39.5 & 43.8 & 50.7 \\
    Number of subtitles & 6,182 & 6,998 & 8,520 \\
    Number of speaker occurrences & 501 & 732 & 951 \\
    \hline
    \hline
    Number of scenes & 206 & 249 & 390 \\
    Proportion of spoken scenes (\%) & 94.7 & 97.2 & 97.2 \\
    \hline
    \hline
    Number of speakers per scene (average) & 2.43 & 2.94 & 2.44 \\
    Number of speakers per scene (standard deviation) & 1.22 & 1.52 & 1.14 \\
    \hline
  \end{tabular}
\end{table}

As can be seen in Table~\ref{tab:corpus}, the duration of the spoken parts in each \textsc{tv} serial covers in average a bit less than half of the total duration of the films. This proportion is expressed as a percentage named \textit{speech coverage} in the table. Speech is more represented in~\textit{House of Cards} (coverage $\simeq 50\%$ of the total time with 8,520 subtitles) than in the other two series.

Speech is uniformly distributed over the scenes, with in average more than $95\%$ of the scenes containing at least one subtitle, which suggests that most social interactions are expressed verbally in these three \textsc{tv} serials.

Furthermore, the average number of speakers by scene remains quite low (ranging from 2.43 to 2.94 depending on the \textsc{tv} series), often resulting in simple patterns of verbal interactions properly handled by applying the basic heuristics described in Subsection~\ref{subsec:seq_estimate}.

We now turn to the evaluation of the accuracy of the four basic rules we use to estimate verbal interactions from the only sequence of speaker-labeled speech turns.

\subsection{Conversational Interactions}
The evaluation of the way we estimate verbal interactions from the sequence of speech turns is performed in two ways. First, \textit{directly}, by measuring the performance of our method for achieving the task of estimating interactions. Second, \textit{indirectly}, by measuring the reliability of the cumulative network (cf. Subsection \ref{sec:cum_net}) resulting from the application of the four basic heuristics introduced in Subsection \ref{subsec:seq_estimate}. We first describe the episode sample we annotated for this part of the evaluation process.

\vspace{0.3cm} \noindent \textbf{Sample of test episodes.}
In order to evaluate the reliability of the methods introduced for estimating verbal interactions, a subset of test episodes for each of the three \textsc{tv} serials is selected. For each series, the considered subset of episodes is defined so that the distribution of the number of speakers per scene remains representative of the same distribution observed in the first 10 episodes of each series.

The most straightforward way of computing the frequency of the number $n$ of speakers per scene consists in computing the proportion of scenes containing $n = 1, 2, ...$ speakers. 
However, by doing so, one ignores the length of the scenes in terms of utterances, which can be different even between two scenes containing the same numbers of speakers.
Using this approach to build our sample could for instance result in the same proportion of three-speaker scenes as in the whole corpus, but with significantly shorter scenes, artificially resulting in fewer complex patterns of speech turns, and in better performances when estimating speaker interactions.
In order to address this issue, we used a different method, by considering the proportion of \textit{utterances} belonging to scenes with $n$ speakers (instead of the proportion of \textit{scenes}).

\begin{figure}[!ht]
  \vspace{2mm}
  \centering
  \includegraphics[width=.49\textwidth]{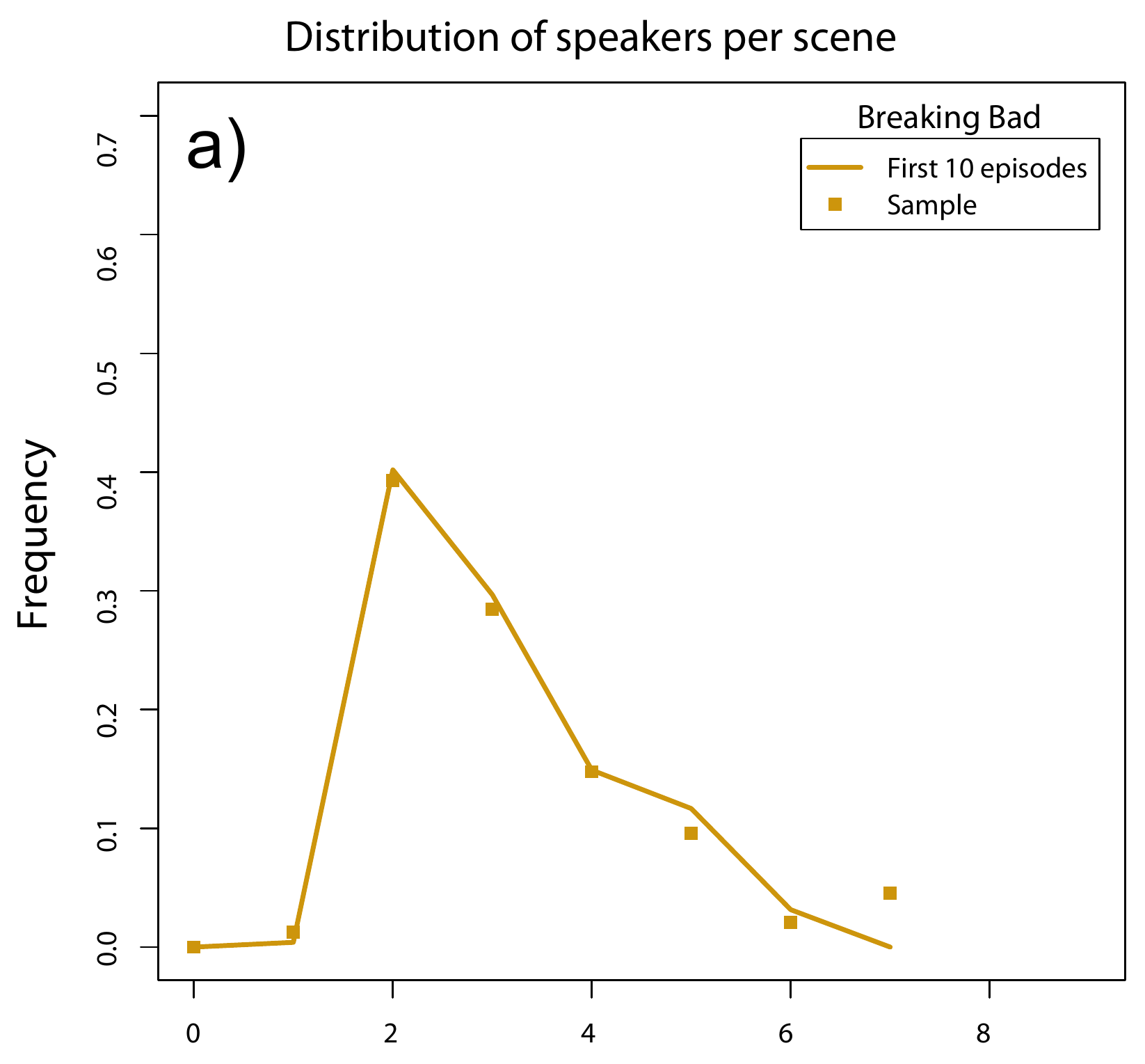}
  \includegraphics[width=.49\textwidth]{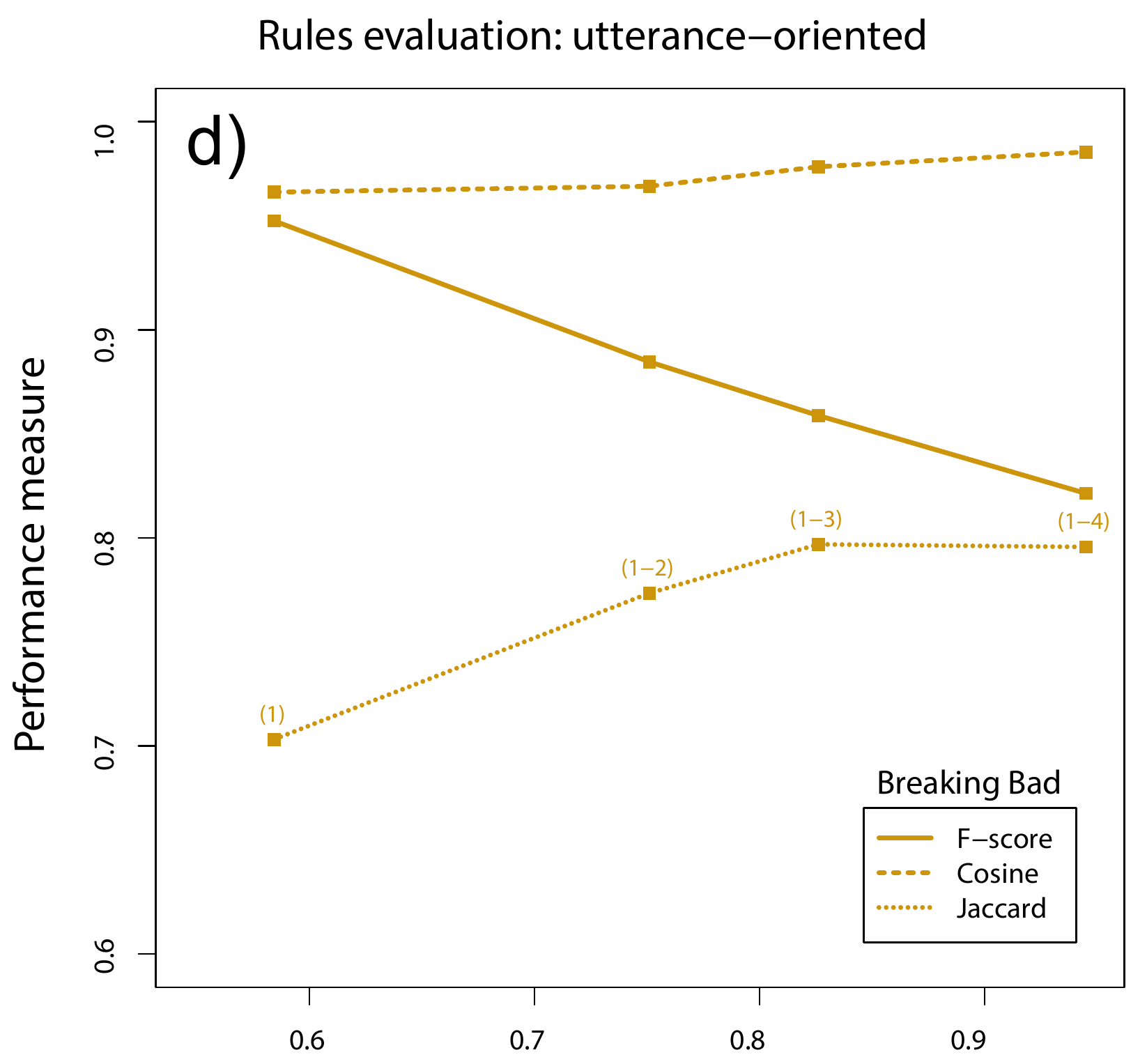}\\
  \includegraphics[width=.49\textwidth]{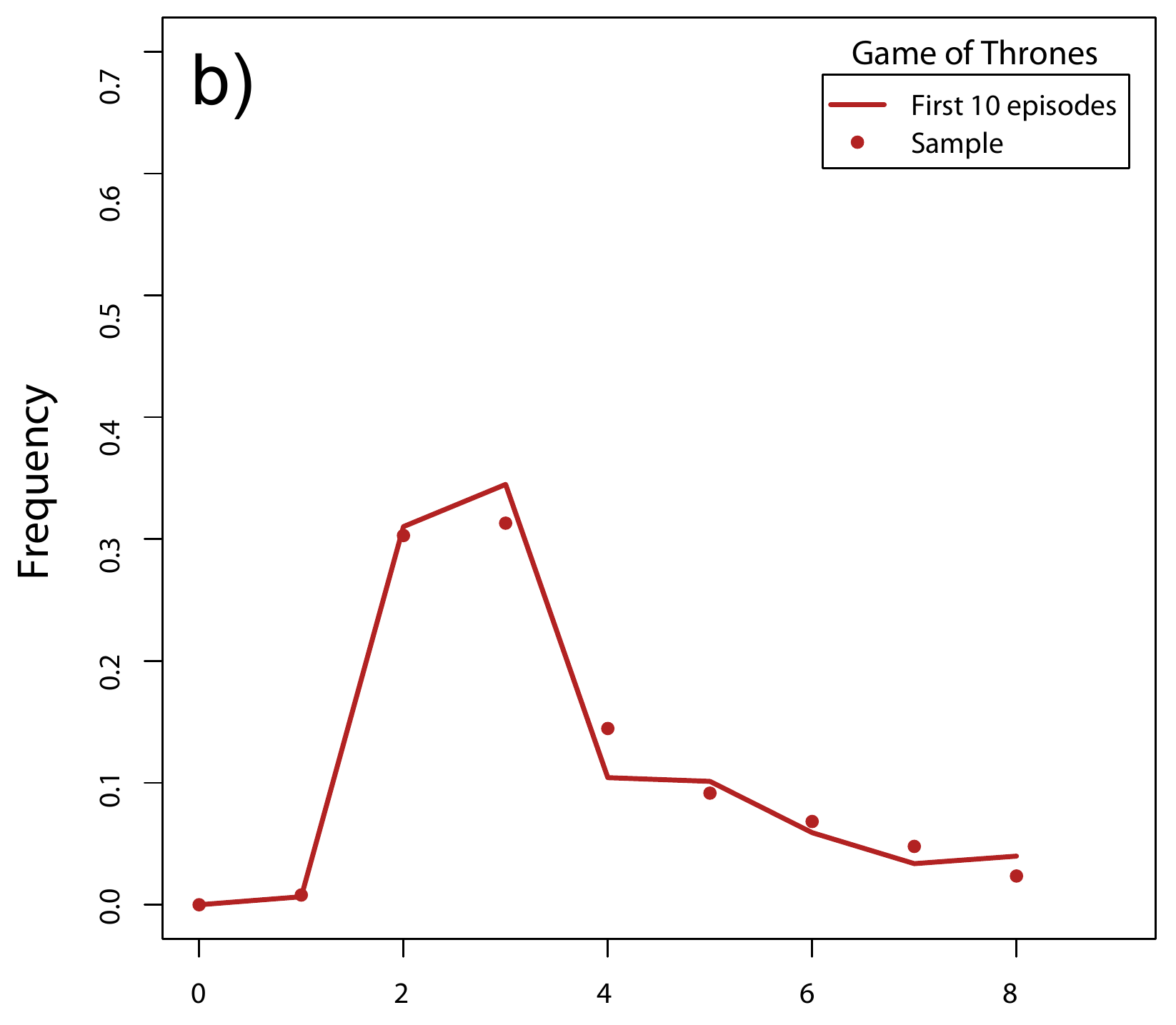}
  \includegraphics[width=.49\textwidth]{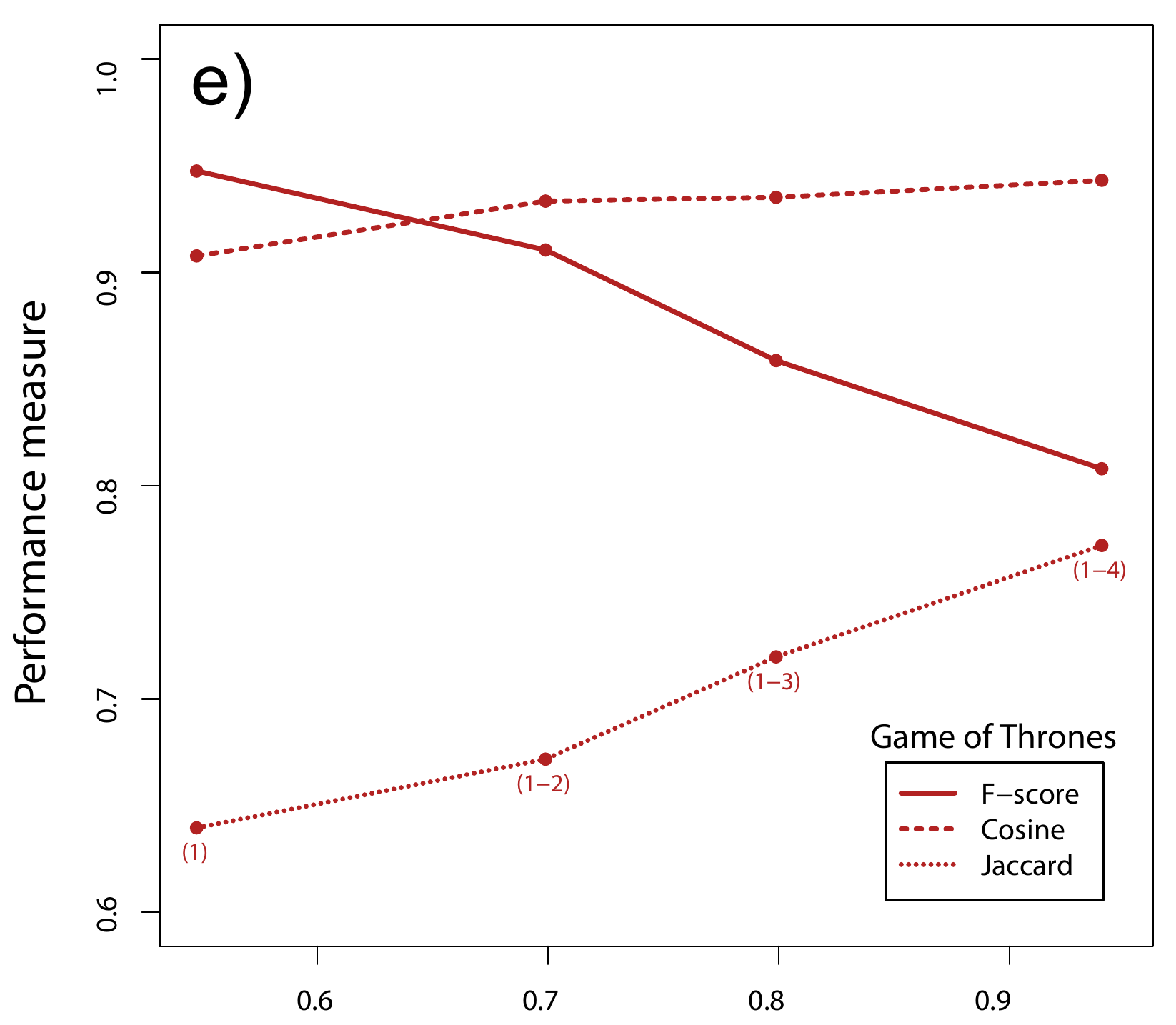}\\
  \includegraphics[width=.49\textwidth]{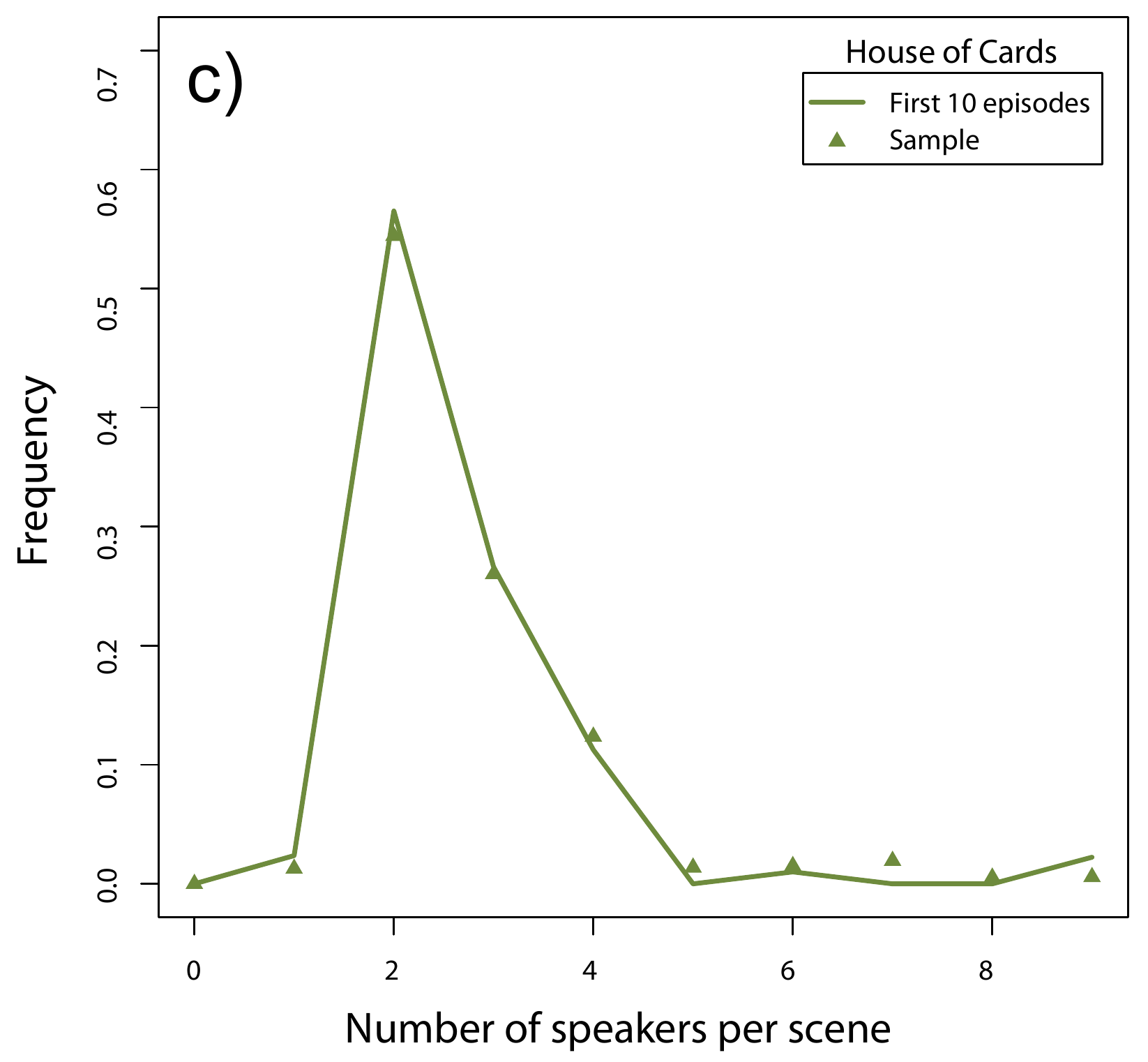}
  \includegraphics[width=.49\textwidth]{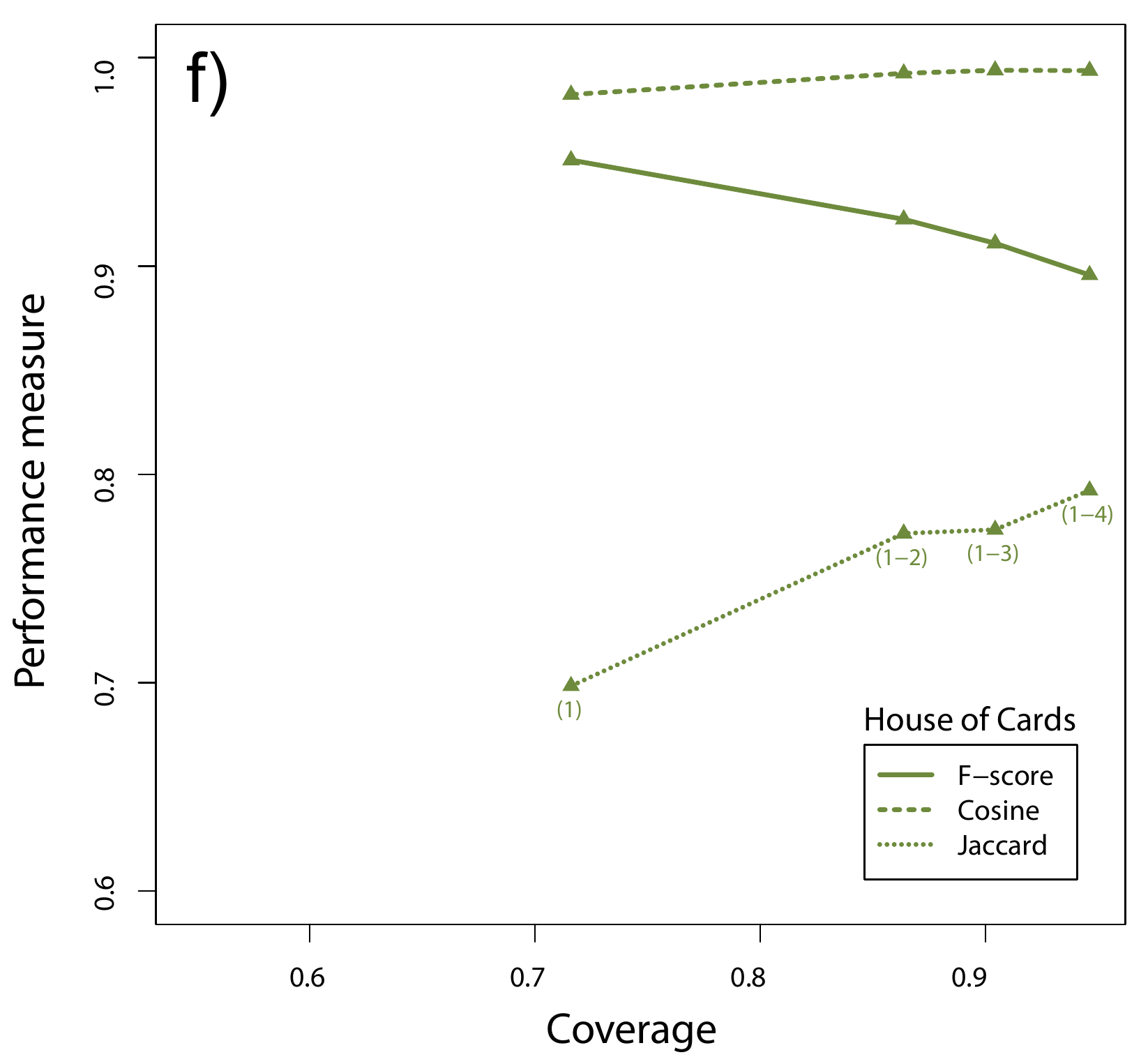}
  \caption{\label{fig:step_by_step_rules_eval} Left column: distribution of speakers per scene, over the first 10 episodes (lines) and over the samples (points). Right column: step-by-step evaluation of the rules used for sequentially estimating verbal interactions. Each row is dedicated to a specific \textsc{TV} series.}

\end{figure}

The plots constituting the left column in Figure~\ref{fig:step_by_step_rules_eval} show the distribution of the number of speakers by scene computed in this way. The lines correspond to the distribution observed in the corpus subset that we introduced in Subsection~\ref{subsec:corpus}, containing about 10 episodes for each of the three \textsc{tv} serials. The points represent the three episode samples we chose as test subsets for each \textsc{tv} serial.

\begin{table}[!ht]
  \caption{\label{tab:test_corpus}{\it Test corpus: main features.}}
  \centering
  \begin{tabular}{|l||c|c|c|}
    \hline
    \textsc{test} subset & \textsc{BB} & \textsc{GoT} & \textsc{HoC} \\
    \hline
    \hline
    Number of episodes & 4 & 3 & 3 \\
    Episodes & 4, 6, 10, 11 & 3, 7, 8 & 1, 7, 11 \\
    Total duration (\textsc{h:mm:ss)} & 2:59:31 & 2:37:17 & 2:29:23 \\
    Speech coverage (\%) & 45.16 & 48.2 & 44.2 \\
    Number of subtitles & 2,254 & 2,282 & 2,194 \\
    Number of speaker occurrences & 202 & 233 & 231 \\
    \hline
    \hline
    Number of scenes & 82 & 81 & 99 \\
    Proportion of spoken scenes (\%) & 95.1 & 97.5 & 97.0 \\
    \hline
    \hline
    Number of speakers per scene (average) & 2.46 & 2.88 & 2.33 \\
    Number of speakers per scene (standard deviation) & 1.14 & 1.49 & 1.11 \\
    \hline
  \end{tabular}
\end{table}

For each of the three resulting \textsc{tv} serial subsets, the same features as those computed in Table~\ref{tab:corpus} are reported in Table~\ref{tab:test_corpus}. For each episode of the three test sets, each utterance is manually labeled according to the speakers it is intended for. For monologues, where no specific listener is targeted, a special \textit{null} label is introduced, and in case of multiple addressees for a single utterance, multiple labels were assigned.

\vspace{0.3cm} \noindent \textbf{Evaluation metrics.}
In estimating verbal interaction, the decision is made at the utterance level, by assigning to each utterance the speaker(s) it is intended for. The task then consists in categorizing every utterance among the available speaker classes, with multiple classes allowed if the utterance is labeled as addressed to multiple characters. Standard performance measures used in Information Retrieval in order to evaluate the multi-label categorization task can therefore be used as direct evaluation metrics. More specifically, we perform this direct evaluation of the basic rules we introduced for estimating verbal interactions by using the following evaluation procedures discussed in \parencite{Tsoumakas2006} for multi-label categorization:
\begin{enumerate}
	\item \textbf{Recall}:
	\begin{equation}
		R(\gamma, \mathbb{X}) = \frac{1}{|\mathbb{X}|} \sum_{y \in\mathbb{X}} \frac{|\gamma(y) \cap M(y)|}{|M(y)|}
	\end{equation}
where $\mathbb{X}$ denotes the utterance set; $\gamma(y)$ denotes the set of interlocutor(s), possibly multiple, hypothesized for the utterance $y$; and $M(y)$ the set of reference interlocutors, as manually labeled, for the utterance $y$. In this context, the Recall is the average proportion, for every utterance $y$, of retrieved interlocutors among the reference ones (i.e. the proportion of false negatives).

\item \textbf{Precision}:
	\begin{equation}
		P(\gamma, \mathbb{X}) = \frac{1}{|\mathbb{X}|} \sum_{y \in\mathbb{X}} \frac{|\gamma(y) \cap M(y)|}{|\gamma(y)|}
	\end{equation}
In this context, the Precision corresponds to the average proportion, for every utterance $y$, of relevant interlocutors among the retrieved ones (i.e. the proportion of false positives).

\item \textbf{$F$-score}:
	\begin{equation}
	F(\gamma, \mathbb{X}) = \frac{2P(\gamma, \mathbb{X})R(\gamma, \mathbb{X})}{P(\gamma, \mathbb{X})+R(\gamma, \mathbb{X})}
	\end{equation}
The $F$-score is the harmonic mean of Precision and Recall, traditionally used in the Information Retrieval domain. Compared to the arithmetic mean, it allows putting more contrast on situations where both the Precision and Recall reach high values.

\end{enumerate}

Besides this direct evaluation of the task of estimating verbal interactions, we also perform an \textit{indirect} evaluation through the assessment of the resulting cumulative conversational network (cf. Subsection \ref{sec:cum_net}). For this purpose, we compare the cumulative network obtained through our method with a cumulative network extracted manually, which constitutes our ground truth. Following the method described in \parencite{Agarwal2013} for a similar purpose, we first convert the adjacency matrix of each one of these two networks into a vector by simple column concatenation. We then measure the similarity between the two resulting vectors. When interactions are weighted, the estimated and ground-truth networks are compared by computing the normalized Euclidean distance and the cosine similarity between the two vectors of edge weights. When focusing on the mere fact that two characters verbally interact with each other whatever the interaction amount, we do not weight interactions, and we evaluate their similarity by computing the Jaccard index of the two sets of edges, both as estimated and as manually labeled. The measures of similarity between the estimated and ground-truth networks are computed both when discarding the first and last utterances of each scene and when considering every speech segment: the first utterance of the next scene is sometimes slightly anticipated at the very end of the current one, possibly resulting in irrelevant interactions.

By using both direct and indirect evaluation metrics, it is possible to measure the performance of the rules we used for sequentially estimating the verbal interactions.

\vspace{0.3cm} \noindent \textbf{Evaluation results.}
In evaluating the performance of the rules used for sequential estimate of verbal interactions, we follow a step-by-step process, by successively using in conjunction to the first, most robust rule, the three remaining ones. The plots on the right side of Figure~\ref{fig:step_by_step_rules_eval} show the changes when applying a more and more comprehensive set of rules, from the single first one, denoted \textit{(1)}, to the whole four rules, denoted \textit{(1--4)}: the changes are expressed in terms of coverage (proportion of utterances processed by applying different subsets of rules), and performance (directly through the $F$-score, and indirectly through the network similarity measures).

Not surprisingly, the more rules are used, the more interactions are hypothesized. As reported in Table~\ref{tab:rules_eval}, the very basic first rule (\textit{surrounded speech turn}) allows in average to hypothesize interlocutors for 62\% of the spoken segments. When the whole set of rules is used, decisions are made for 94\% of the utterances. The remaining utterances correspond to soliloquies or isolated utterances.

\begin{table}[!ht]
  \caption{\label{tab:rules_eval}{\it Evaluation of the joint use of the rules applied for sequentially estimating speakers interactions. The cosine, L2, Jaccard measures are both computed when discarding the first and last segments of each scene, or not.}}
  \centering
  \begin{tabular}{|c|c|c!{-}c|c!{-}c|c!{-}c||c!{-}c|}
    \hline
    \multirow{2}*{Rules} & \multirow{2}*{Evaluation metrics} & \multicolumn{8}{c|}{\textsc{tv} serial} \\
    \cline{3-10}
    & & \multicolumn{2}{c}{\textsc{bb}} & \multicolumn{2}{c}{\textsc{got}} & \multicolumn{2}{c||}{\textsc{hoc}} & \multicolumn{2}{c|}{\textbf{Average}} \\
    \hline
    \hline
    \multirow{7}*{\textit{(1)}} & \textbf{Coverage} & \multicolumn{2}{c|}{0.58} & \multicolumn{2}{c|}{0.55} & \multicolumn{2}{c||}{0.72} & \multicolumn{2}{c|}{0.62} \\
    \cline{2-10}
    & \textbf{$F$-score} & \multicolumn{2}{c|}{0.95} & \multicolumn{2}{c|}{0.95} & \multicolumn{2}{c||}{0.95} & \multicolumn{2}{c|}{0.95} \\
    & \textbf{Precision} & \multicolumn{2}{c|}{0.96} & \multicolumn{2}{c|}{0.95} & \multicolumn{2}{c||}{0.95} & \multicolumn{2}{c|}{0.95} \\
    & \textbf{Recall} & \multicolumn{2}{c|}{0.94} & \multicolumn{2}{c|}{0.94} & \multicolumn{2}{c||}{0.95} & \multicolumn{2}{c|}{0.94} \\
    \cline{2-10}
    & \textbf{Jaccard similarity} & \multicolumn{2}{c|}{0.70} & \multicolumn{2}{c|}{0.64} & \multicolumn{2}{c||}{0.70} & \multicolumn{2}{c|}{0.68} \\
    & \textbf{Cosine similarity} & 0.97 & 0.96 & 0.91 & 0.89 & 0.99 & 0.99 & 0.96 & 0.95 \\
    & \textbf{L2 distance} & 0.26 & 0.29 & 0.43 & 0.47 & 0.19 & 0.16 & 0.29 & 0.31 \\
    \hline
    \hline
    \multirow{7}*{\textit{(1--4)}} & \textbf{Coverage} & \multicolumn{2}{c|}{0.94} & \multicolumn{2}{c|}{0.94} & \multicolumn{2}{c||}{0.95} & \multicolumn{2}{c|}{0.94} \\
    \cline{2-10}
    & \textbf{$F$-score} & \multicolumn{2}{c|}{0.82} & \multicolumn{2}{c|}{0.81} & \multicolumn{2}{c||}{0.90} & \multicolumn{2}{c|}{0.84} \\
    & \textbf{Precision} & \multicolumn{2}{c|}{0.84} & \multicolumn{2}{c|}{0.82} & \multicolumn{2}{c||}{0.90} & \multicolumn{2}{c|}{0.85} \\
    & \textbf{Recall} & \multicolumn{2}{c|}{0.80} & \multicolumn{2}{c|}{0.80} & \multicolumn{2}{c||}{0.89} & \multicolumn{2}{c|}{0.83} \\
    \cline{2-10}
    & \textbf{Jaccard similarity} & \multicolumn{2}{c|}{0.80} & \multicolumn{2}{c|}{0.77} & \multicolumn{2}{c||}{0.79} & \multicolumn{2}{c|}{0.79} \\
    & \textbf{Cosine similarity} & 0.99 & 0.98 & 0.94 & 0.93 & 0.99 & 0.99 & 0.97 & 0.97 \\
    & \textbf{L2 distance} & 0.17 & 0.20 & 0.34 & 0.37 & 0.11 & 0.13 & 0.21 & 0.23 \\
    \hline
    \end{tabular}
\end{table}

More surprisingly, as can be seen both on the right-and plots of Figure~\ref{fig:step_by_step_rules_eval} and in Table~\ref{tab:rules_eval}, the additional rules~\textit{(2--4)} introduce more and more mistakes when hypothesizing interlocutors at the utterance level, resulting in a lower and lower $F$-score, but in the meantime, the indirect evaluation measures of the resulting network are improving: as can be seen in Table~\ref{tab:rules_eval}, while the $F$-score decreases in average from 0.95 to 0.84, the Euclidean distance between the estimated and the ground-truth networks decreases from 0.29 to 0.21 if the first and final utterances of every scene are discarded, or from 0.31 to 0.23 if not.

Such a discrepancy suggests that errors made locally when assigning each utterance to the addressed characters do not deteriorate the reliability of the resulting conversational network. Indeed, only a small proportion of the errors made at such a local level (utterance-level $F$-score amounting to zero) introduces irrelevant links in the resulting cumulative network (14.08\% for \textit{Breaking Bad}, 13.43\% for \textit{Game of Thrones} and 5.34\% for \textit{House of Cards}). Moreover, some errors made at the utterance-level by using more and more covering rules allow to retrieve interactions that would otherwise have been missed, or improperly weighted, by carefully applying the only rule~\textit{(1)}. The additional rules~\textit{(2--4)} tend to introduce correct interactions, but at wrong places, and finally result in more reliable conversational networks, with more actual relationships captured and more representative link weights when measuring interaction intensity. In other words, the errors consisting in misplacing an interaction in time does not affect the cumulative network, in which time is integrated. Though basic, the four heuristics introduced in Subsection~\ref{subsec:seq_estimate} turn out to be very effective when building such cumulative conversational networks. 

We now qualitatively evaluate narrative smoothing, the method we use to build the dynamic conversational network from the interactions between speakers.

\subsection{Narrative smoothing}
\noindent \textbf{The protagonists.}
We first base our analysis on the protagonists of the considered \textsc{tv} serials, i.e. the nodes in the corresponding extracted social networks. We present only a small number of results, which concern characters of particular interest. We characterize them using the \textit{node strength}, a generalization of the concept of \textit{node degree} defined in Graph Theory as the sum of the weights of the links attached to the considered node. In our case, weights are based on spoken interaction durations, so the strength of a character is related to how much and how frequently he speaks to others.

\begin{figure}[!ht]
  \vspace{2mm}
		\includegraphics[width=.49\linewidth]{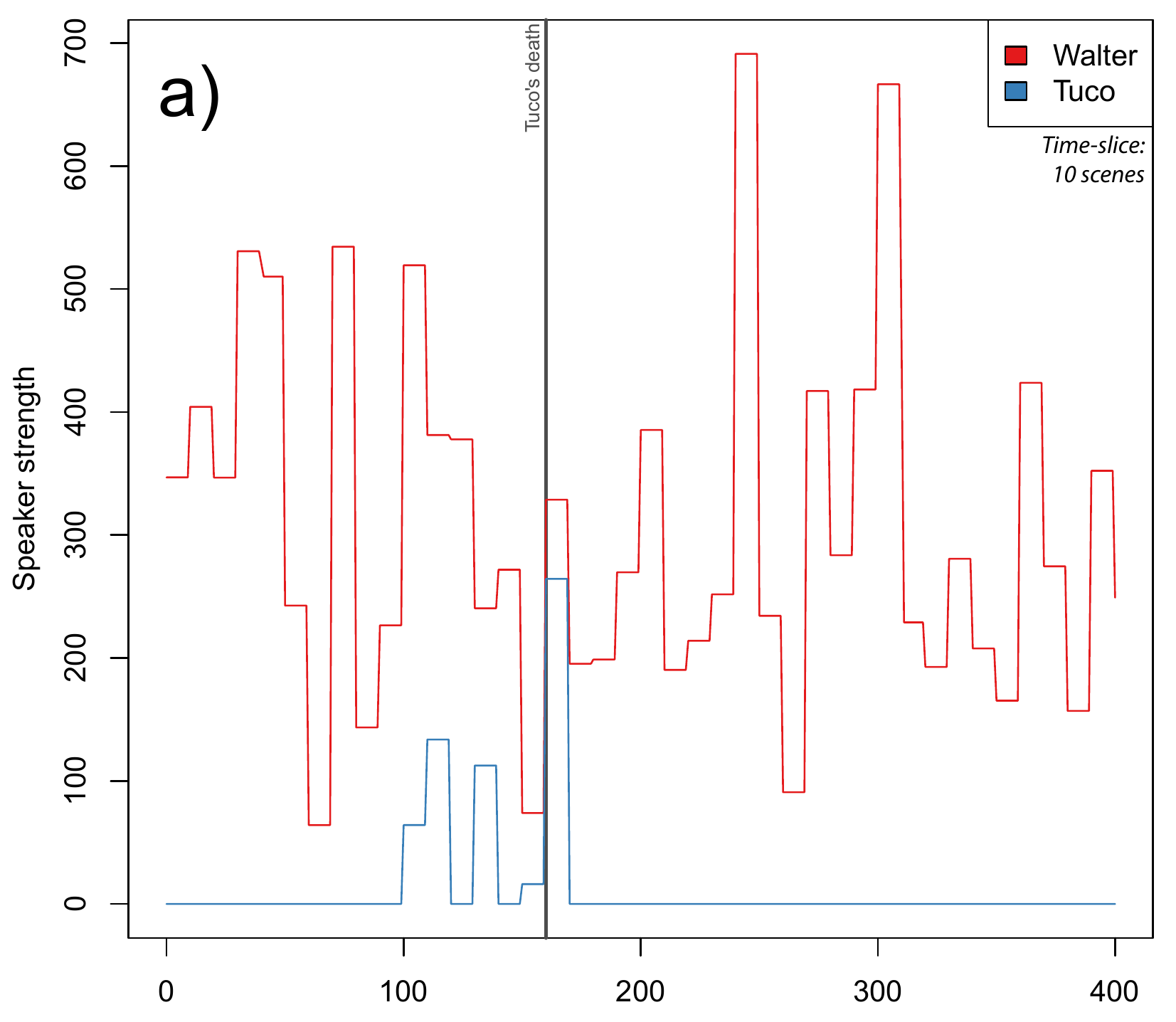}
		\includegraphics[width=.49\textwidth]{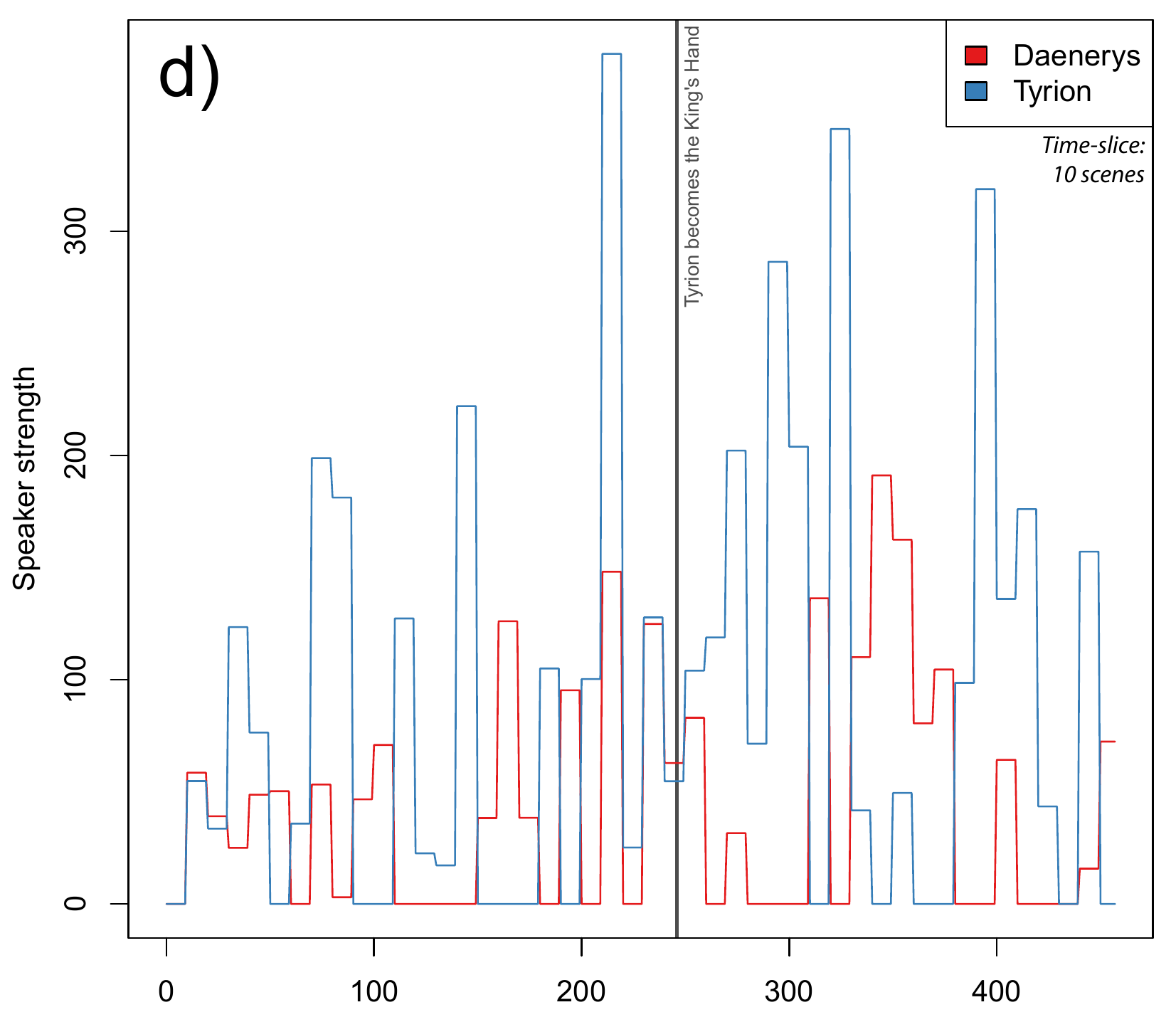}\\
		\includegraphics[width=.49\linewidth]{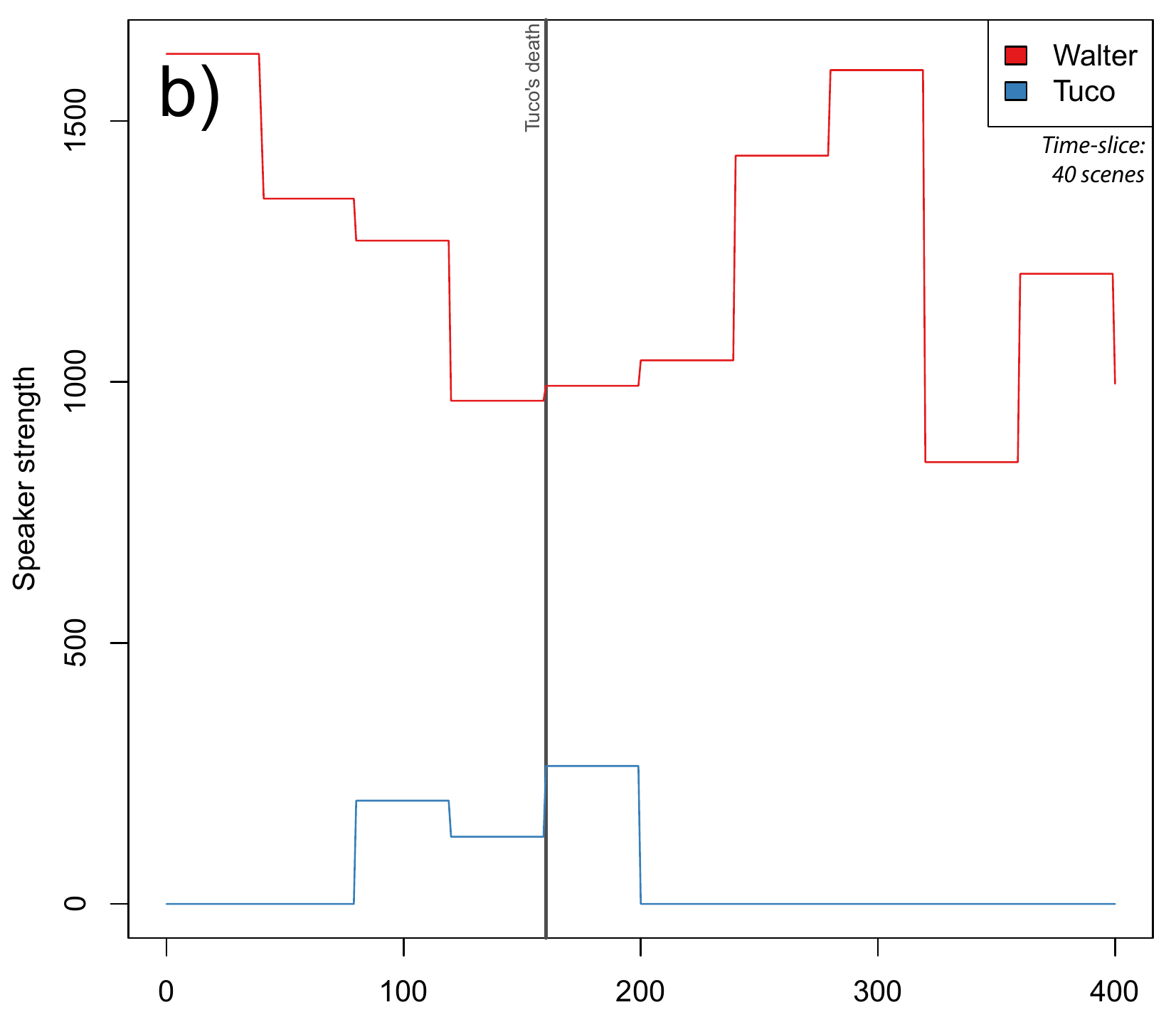}
		\includegraphics[width=.49\textwidth]{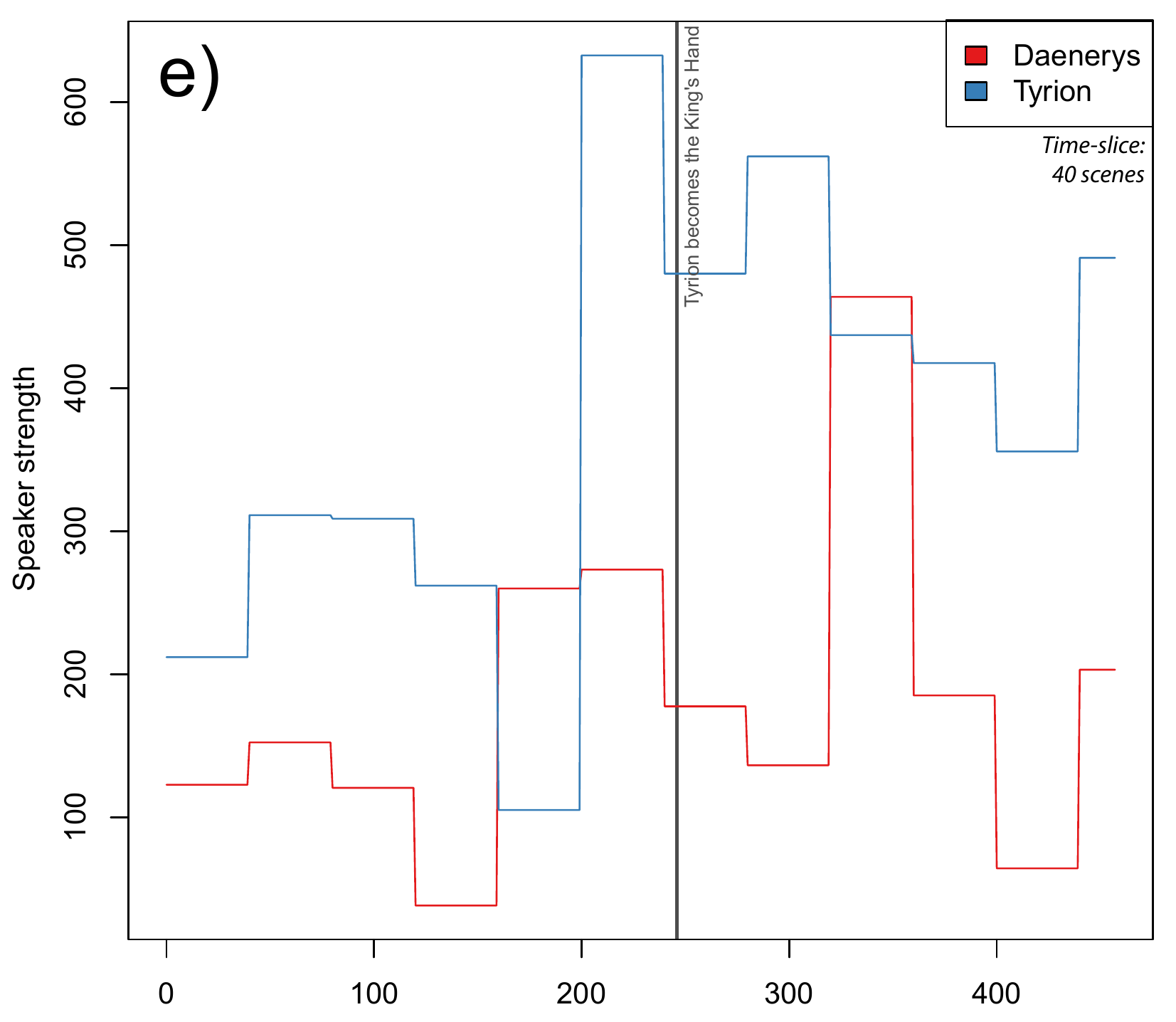}\\
		\includegraphics[width=.49\linewidth]{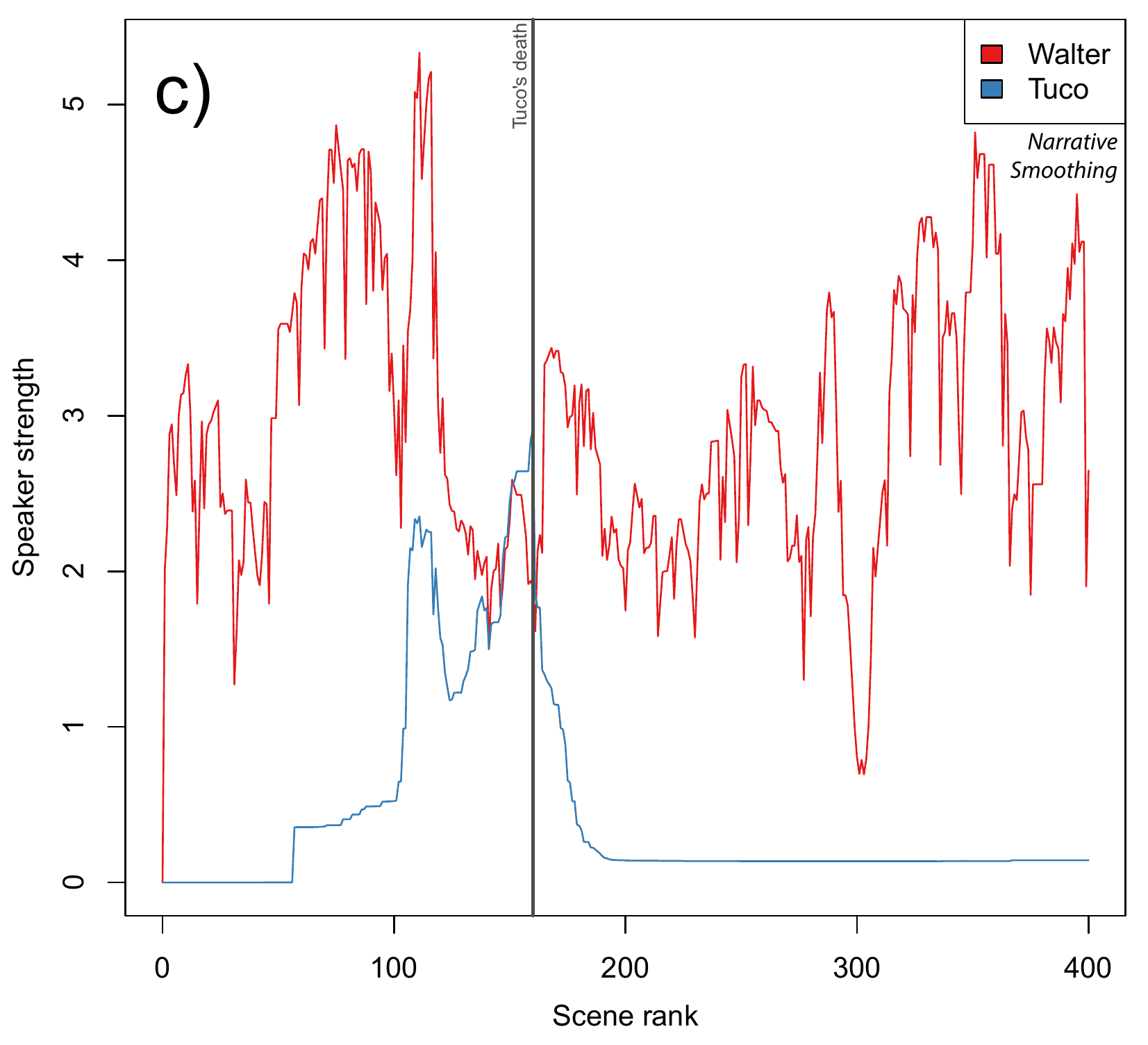}
		\includegraphics[width=.49\textwidth]{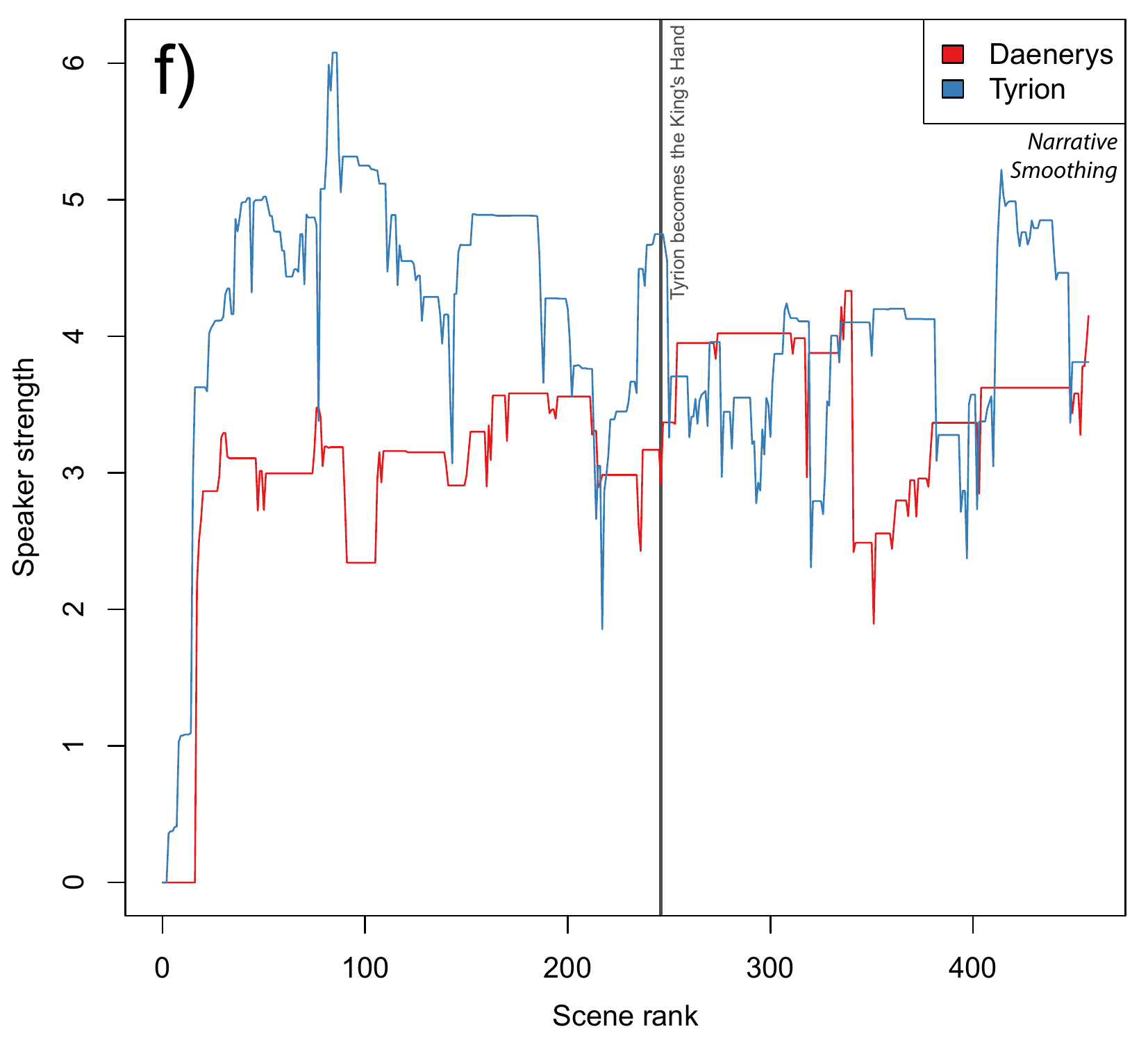}
  \caption{\label{fig:dyn_strength} Strengths of two major characters of \emph{Breaking Bad} (left column) and \emph{Game of Thrones} (right column) plotted as functions of the chronologically ordered scenes. The first and second rows correspond to $10$ and $40$ scenes time-slices, respectively, whereas the bottom row shows the results of narrative smoothing.}
\end{figure}

We first focus on Walter White, the main character of \textit{Breaking Bad}, and Tuco Salamanca, one of the drug dealers with whom he is in business. When considering the cumulative network of \textit{Breaking Bad}, i.e. the temporal integration over the first $20$ episodes, the strength of Walter White (his total interaction time with others) is about twenty times as large as the strength of Tuco: $12,332$ seconds for Walter (rank~1) \textit{vs.} $590$ for Tuco (rank~11). By comparison, the left-hand plots of Figure~\ref{fig:dyn_strength} (plots a--c) display the evolution of their strengths, as a function of time (expressed in terms of scenes). Plots a) and b) were obtained through the use of fixed-size observation windows, set to $10$ scenes (around half an episode) and $40$ scenes (about two episodes), respectively. Plot c) relies on our narrative smoothing method, so it shows the instantaneous strengths. The plot based on the $40$-scene windows (plot b) is consistent with the observation we made on the cumulative graphs, i.e. it shows Walter as much more important than Tuco, at any time. It is also the case with the $10$-scene windows plot (plot a), but to a lesser extent. In particular, Tuco's strength almost reaches that of Walter around scene 165. But the narrative smoothing plot brings a completely different vision of Tuco's role in the story. From scene 100, Tuco's importance tends to increase and even overcomes the importance of the main protagonist for some time, before suddenly decreasing and reaching almost zero after scene 200. This clearly corresponds to a subplot, or a short narrative episode, ending with Tuco's death, at the end of scene 167 (represented as a vertical line on plots a--c).

We now switch to Daenerys Targaryen and Tyrion Lannister, two major protagonists of \textit{Game of Thrones}. The right-hand plots of Figure~\ref{fig:dyn_strength} (plots d--e) show how their strengths evolve over the first two seasons of the series, again as a function of the chronologically ordered scenes, and illustrate the limitations of time-windowing approaches. The appearance of Daenerys' storyline onscreen has a relatively slow pace in these seasons (Figure~\ref{fig:narr_freq}) and as can be seen, when the window is too narrow, this creates noisy, irrelevant measurements of her narrative importance (plot d) in Figure~\ref{fig:dyn_strength}). It appears very unstable because her storyline alternates with many others on the screen. A wider observation window (plot e) on the same figure) is more likely to cover successive occurrences of Daenerys in the narrative, but, unlike our narrative smoothing method, prevents us from locating precisely the scenes responsible for Tyrion's current importance. For instance, a local maximum in Tyrion's strength is reached at scene 247 (plot d) in Figure~\ref{fig:dyn_strength}), just after a major narrative event took place~--~the nomination of Tyrion as the King's Counselor (represented as a vertical line in plots d--e). Such an event remains unnoticed when accumulating the interactions during too large time-slices (plot e) in Figure~\ref{fig:dyn_strength}), but is well captured by our approach.

Figure~\ref{fig:dyn_strength} also reveals an important property of our way of building the dynamic network. Because the past (resp. future) occurrences of a particular relationship are still (resp. already) active as long as the involved characters do no interact with others in the meantime, the respective strengths of the main characters of the story appear remarkably balanced. Although Tyrion looks much more central than Daenerys in the time-slice based dynamic networks, whatever the size of the observation window, Daenerys is nearly as central as Tyrion in the network based on our narrative smoothing method: few of her acquaintances are shown onscreen as interacting with others. On the opposite, the story focuses more frequently on Tyrion, but also on separate interactions of his usual interlocutors, weakening his instantaneous strength (especially after scene 252): the dynamic strength, as computed after applying narrative smoothing, does not reduce to a global centrality measure, but also corresponds to a more local property, that measures how exclusively a character is related to his/her social neighborhood.

Our results confirm that cumulative networks, by neglecting the temporal dimension, tend to completely miss punctual changes in the importance of certain characters relatively to the plot. The time-slice based methods can handle the network dynamics, however our observations illustrate that they cannot properly tackle the narrative issue we described in Subsection~\ref{subsec:time_slice}. The choice of an appropriate time window is a particularly sensitive point. By comparison, narrative smoothing captures the state of a relationship at any moment of the plot, using a time scale which directly depends on the narrative pace of the considered series. This allows to finely evaluate the degree of instantaneous involvement of any character in the plot.

\vspace{0.3cm} \noindent \textbf{The relationships.}
We now consider relationships between pairs of characters, instead of single individuals. We characterize each relation depending on its weight, i.e. the amount of time the characters talked to each other, either cumulated over time-slices, possibly consisting of the whole set of episodes, or smoothed with respect to the narrative. Like for the protagonists, we focus on relationships of particular interest.

Let us consider two relationships in~\textit{House of Cards}, representative of two substories: the first one corresponds to a narrative sequence in the storyline related to the main character Francis Underwood~--~his fight with a former ally, the unionist Martin Spinella; the second one is a similar subplot, but related to a secondary character, not as frequently present in the narrative, the journalist Lucas Goodwin, who requests the help of the hacker Gavin Orsay to investigate on Francis. Though locally important in these two substories, neither of these relationships lasts long enough to be noticed in the cumulative network, as resulting from the first two seasons of the series: the interaction time amounts to $562$ seconds for the relation between Francis and Martin, and to $294$ seconds for the relation between Gavin and Lucas. These total interaction times remain quite small compared to the central relation between Francis and his wife Claire, amounting to $2,319$ seconds.

\begin{figure}[!ht]
	\vspace{2mm}
	\includegraphics[width=.49\textwidth]{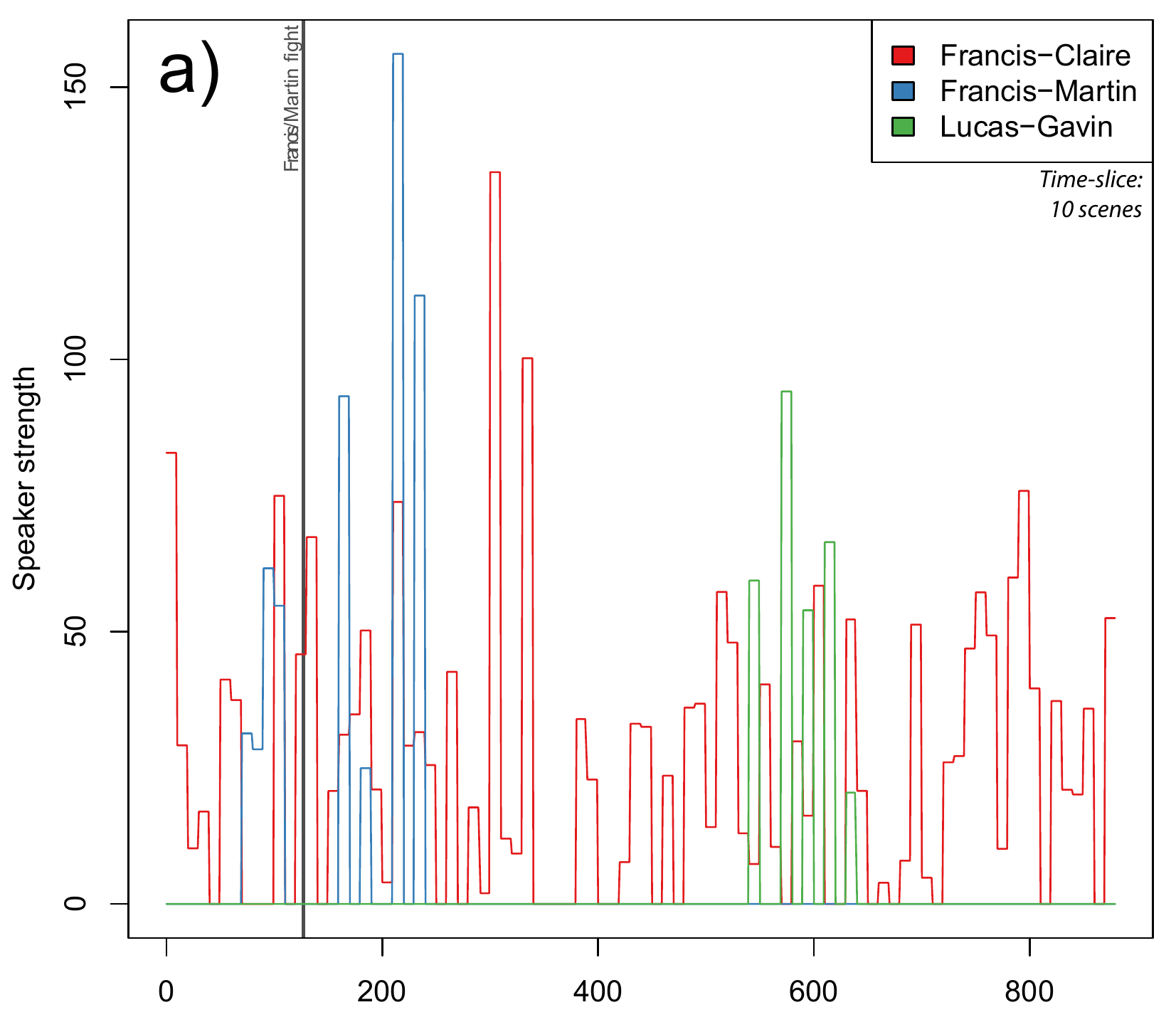}
	\includegraphics[width=.49\textwidth]{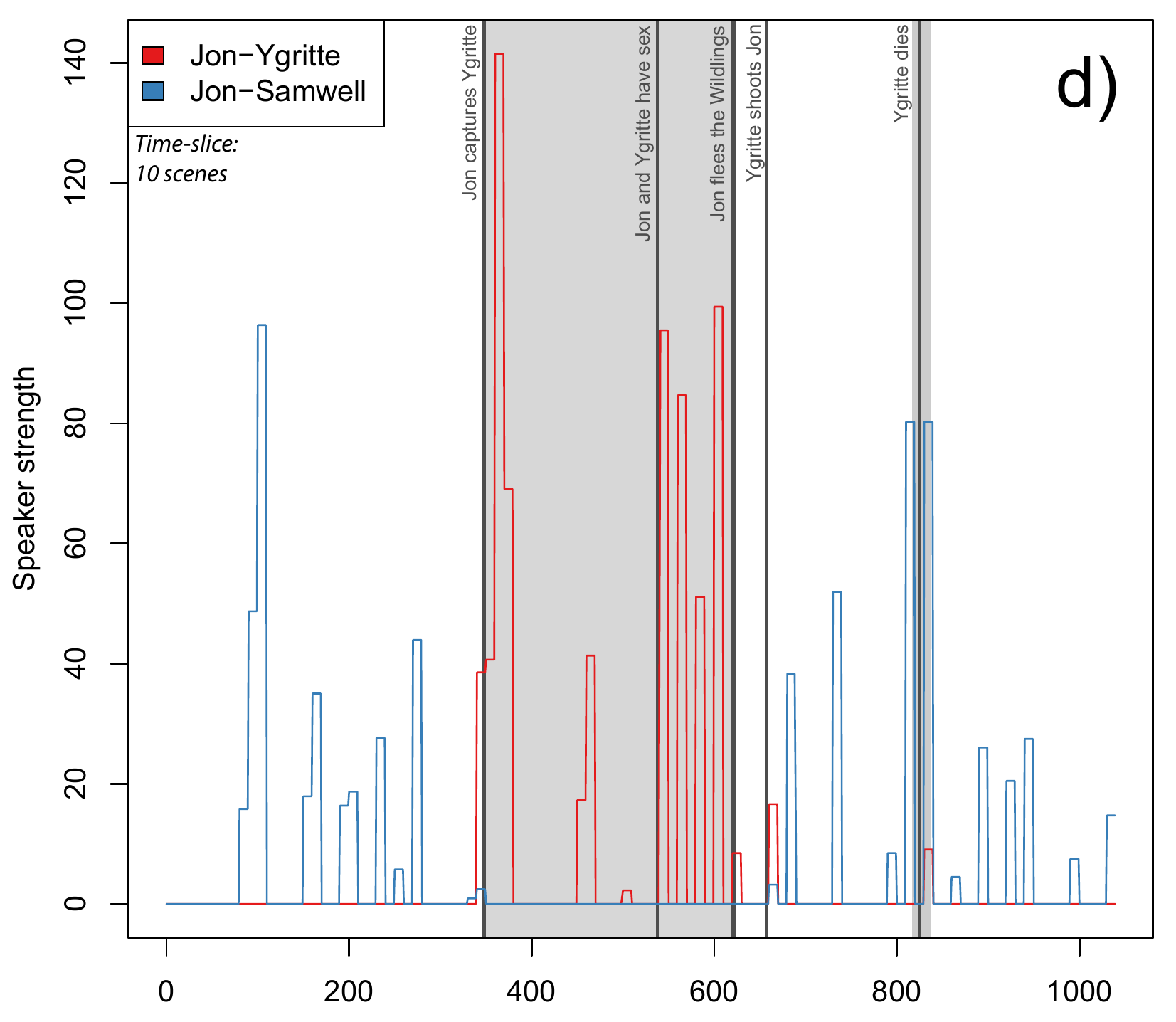}\\
	\includegraphics[width=.49\textwidth]{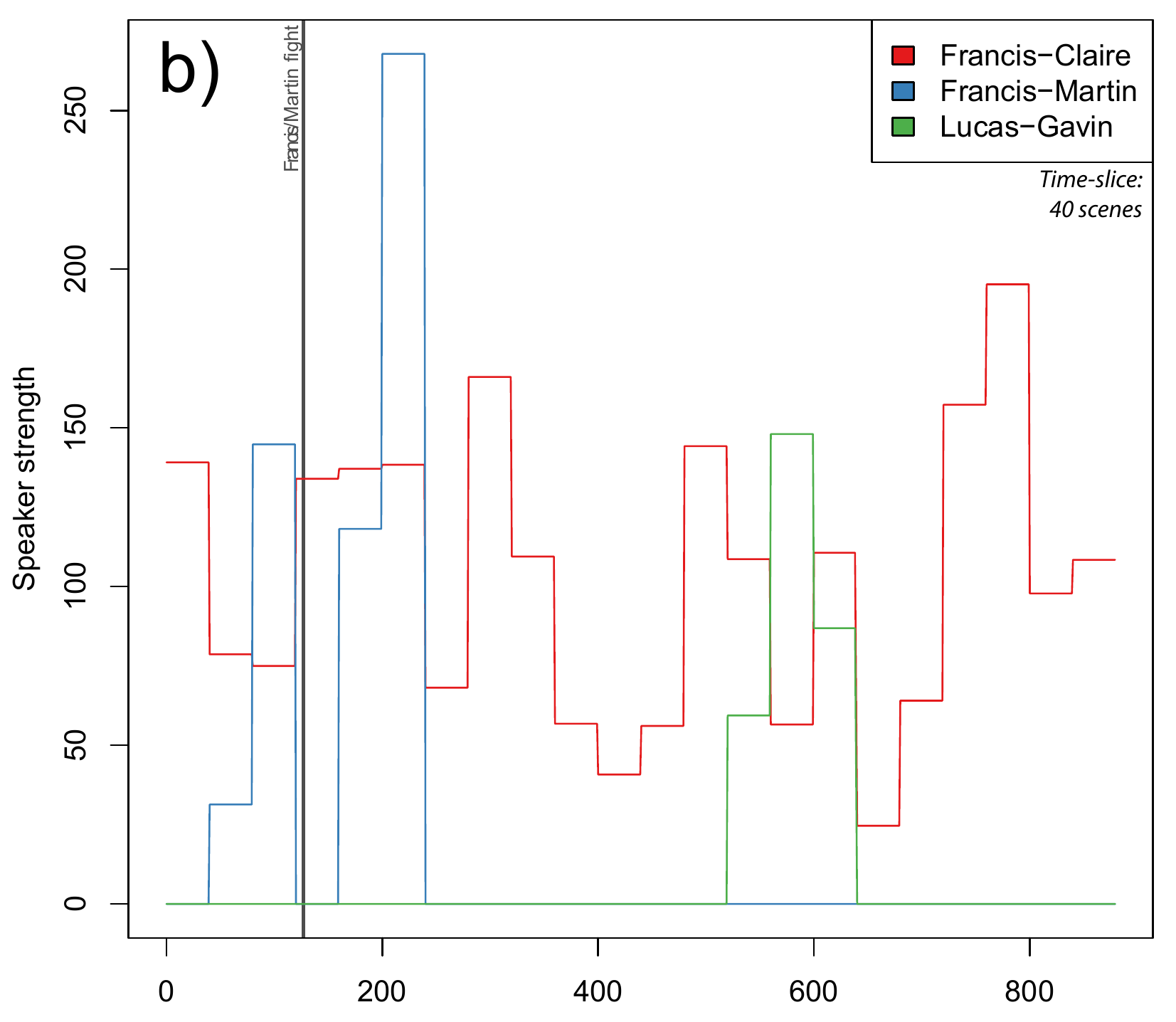}
	\includegraphics[width=.49\textwidth]{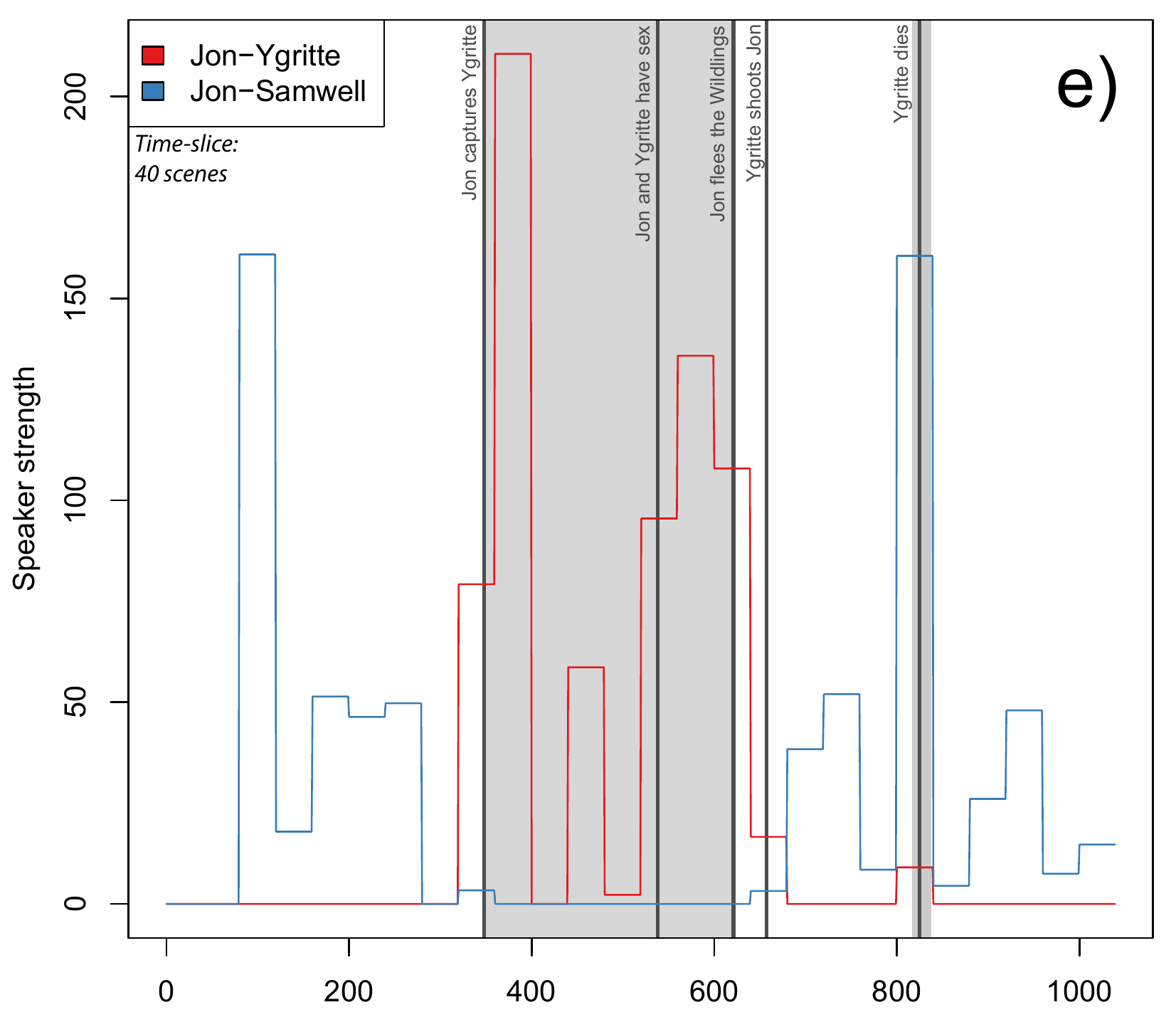}\\
	\includegraphics[width=.49\textwidth]{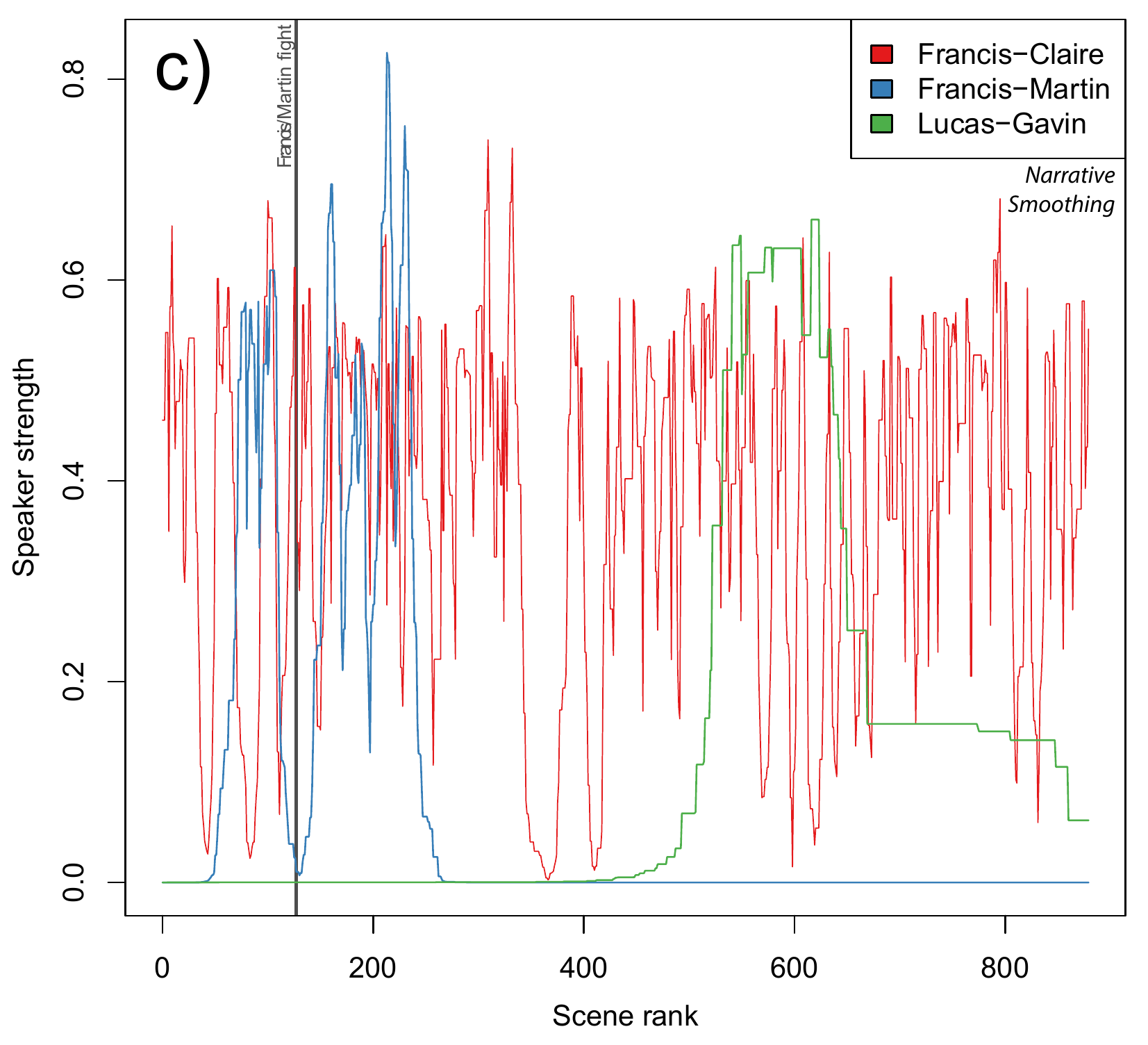}
	\includegraphics[width=.49\textwidth]{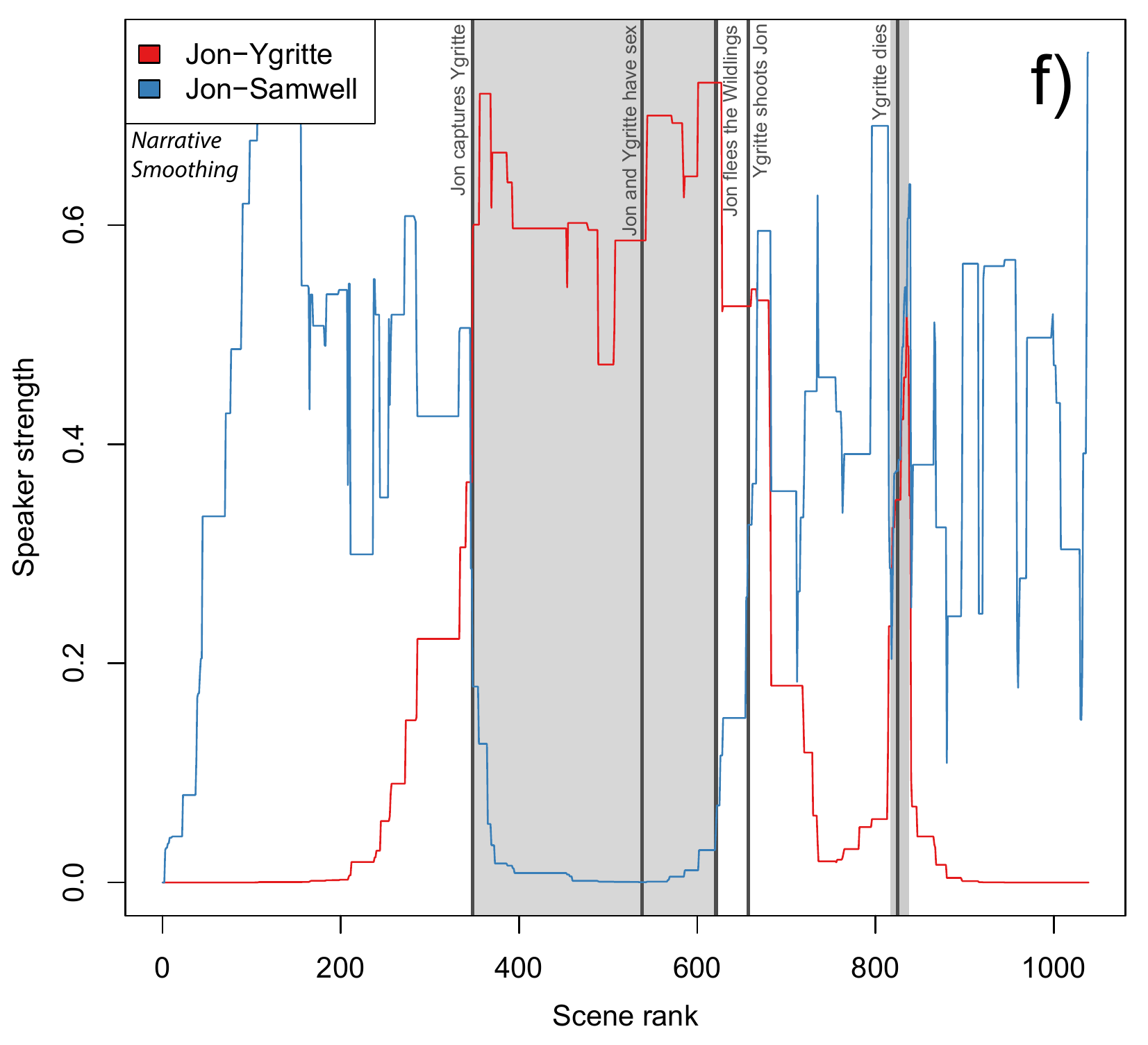}
	\caption{\label{fig:dyn_weight} Weights of several major relationships of \emph{House of Cards} (left column) and \emph{Game of Thrones} (right column) plotted as functions of the chronologically ordered scenes. The first and second rows correspond to $10$ and $40$ scenes time-slices, respectively, whereas the bottom row shows the results of narrative smoothing.}
\end{figure}

Nonetheless, once plotted as a function of the chronologically ordered scenes, as shown in the left-hand plots of Figure~\ref{fig:dyn_weight} (plots a--c), the respective weights of these relationships in the narrative look quite different, whatever the weighting scheme. Both substories, the one based on the relation between Francis and Martin and the one based to the relation between Lucas and Gavin, turn out to be locally as important as the long-term substory based on the relation between the two main characters Claire and Francis. However, all three ways of monitoring these relationships over time are not equivalent. Agglomerating the interactions within short time-slices (plot a) in Figure~\ref{fig:dyn_weight}) makes us miss the continuity of Lucas/Gavin's substory, which occurs \textit{in the narrative} at a slower rate than the substories related to Francis. Conversely, large time-slices (plot b) in Figure~\ref{fig:dyn_weight}) allow to capture this substory, but agglomerate the two main stages of the relation Francis/Martin: before becoming an enemy, Martin is first an ally of Francis. These two parts in the relation correspond to well-separated stages in the narrative, that too large time-slices tend to merge. In contrast, such a breakpoint (materialized in plots a--c by a vertical line located at scene 129) is correctly captured when monitoring the relationship with narrative smoothing.

We now switch back to Game of Thrones, and focus on two links: on the one hand, the romantic relation between Jon Snow and Ygritte, a Wildling, and on the other hand, the friendship between Jon and Samwell Tarly, who also serves in the Night's Watch. Both are represented in the right-hand plots of Figure~\ref{fig:dyn_weight} (plots d--f), with five important events of the Jon-Ygritte relationship, marked by vertical lines on the plots: 1) their first encounter, when Jon captures Ygritte; 2) the moment when they have sex; 3) their separation, when Jon escapes the Wildlings; 4) the vengeance of Ygritte, when she shoots Jon with her bow and arrows; and 5) the death of Ygritte, during the Wildling attack of Castle Black. The first grayed area represents the duration of the romantic relationship between Jon and Ygritte, and the second is the battle of Castle Black. Samwell and Ygritte belong to two separate social groups, with which Jon alternatively interacts: Samwell when he is at Castle Black with the Night's Watch, and Ygritte when he is north of the wall with the Wildlings. This separation very clearly appears in the plot generated through narrative smoothing (plot f). One can distinguish periods of exclusive relationships: with Sam before the capture of Ygritte (first mark), with Ygritte during their romantic relationship (first grayed area); but also periods where Jon interacts with both, such as the Battle of Castle Black (second grayed area). Our method also allows detecting important events, corresponding to peaks of link weight, such as the first time Ygritte and Jon have sex, or the Battle of Castle Black. By comparison, and as noted before, both methods based on time windows fail to show the continuity of these relationships, which appear to be very sporadic in plots d) and e). This is particularly true of the Jon-Ygritte romantic relationship, which appears as completely discontinuous. Of course, this irregularity also hides important events, which do not stand out among these large fluctuations of link weight. Moreover, certain important events are just not associated to important weights, such as the death of Ygritte (rightmost vertical mark).

Our results confirm that cumulative networks are not appropriate for capturing punctual substories supported by specific relationships. Moreover, though much more appropriate for such a task, time-slice approaches suffer from a major drawback: once fixed, the time slice cannot adapt to the variable rates at which the substories appear in the narrative. By overcoming the narrative contingencies, our narrative smoothing approach allows to monitor more accurately over time any relationship, whatever the way the narrative focuses on it. We could confirm its relevance empirically in a separate study \parencite{Bost2016a,Bost2017}. We generated extractive video summaries based on the networks obtained with narrative smoothing, and evaluated them through a user study during which a large sample of viewers were asked to grade them. The clips produced by a hybrid methods combining features extracted from our narratively smoothed conversational networks with lower level multimedia features obtained the best results.

\section{Conclusion and Perspectives}
\label{sec:conclu}
In this paper, we described a novel way of monitoring over time the state of the relationships between characters involved in the usually complex plots of modern \textsc{tv} series. The two methods previously used for this purpose are the cumulative approach, consisting in integrating every relation over the whole considered period of time, and the time-slice approach, consisting in breaking down the time-line into smaller discrete chunks. The first one turns out to be relatively inefficient for investigating complex storylines and a dynamic perspective is more appropriate. The second one complies with this constraint, but defining an appropriate size for the observation window is a very difficult task and constitutes a major drawback: the plots of modern \textsc{tv} series usually consist in parallel storylines shown sequentially onscreen at an unpredictable frequency. As a main consequence, the narrative disappearance in the current scene of some past relationship can usually not be interpreted as a real disappearance, which invalidates the time-slice approach.

To address this issue, we chose to smooth the narrative sequentiality, by considering that the relation between interacting speakers remains active as long as neither of them speaks with others; if so, such separate interactions result in a progressive dissolution of the past link. Symmetrically, the imminence of the next occurrence of the relationship has to increase the link weight.

We then evaluated on our corpus the rules we use for estimating the interacting speakers from the sequence of speaker-labeled speech turns: though possibly misleading punctually, they result in quite reliable estimates of the characters' relationships. We finally experimentally compared our way of building the dynamic network of interacting speakers, which we call \textit{narrative smoothing}, to both mentioned approaches on the three \textsc{tv} serials of our corpus. Though exploratory and qualitative, our results show that our method leads to more relevant results than both other methods, when it comes to instantaneously monitoring the importance of a particular character or of a specific relationship at some point of the story.

The way some characters temporarily aggregate at some point of the story in a community-like structure suggests some narrative sequences result in the stabilization, possibly temporarily, of certain areas in the network. By automatically detecting such a narrative stabilization of some groups of relationships, it should be possible to split the whole story into sub-stories, without assuming a static, predefined, community structure. Finally, the statistical properties of such a dynamic network, as based on the smoothing of the narrative, have still to be studied: the relative balance between the important characters suggests, for instance, that the traditional heavy-tailed degree distribution may not stand in this case.

\section*{Acknowledgments}
\addcontentsline{toc}{section}{Acknowledgments}
This work was supported by the French National Research Agency (ANR) GAFES project (ANR-14-CE24-0022) and the Research Federation Agorantic, University of Avignon.

\renewcommand*{\bibfont}{\footnotesize}
\addcontentsline{toc}{section}{\refname}
\printbibliography

\end{document}